\documentclass[fleqn,usenatbib]{mnras}

\usepackage{mathptmx}

\usepackage[T1]{fontenc}
\usepackage{ae,aecompl}

\usepackage{savesym}
\usepackage{amsmath}
\savesymbol{iint}
\usepackage{txfonts}
\restoresymbol{TXF}{iint}

\usepackage{graphicx}	
\usepackage{amssymb}	
\usepackage[ansinew]{inputenc}
\usepackage{amsfonts}
\usepackage{graphicx}
\usepackage{epsfig}
\usepackage{color}
\usepackage[normalem]{ulem}
\usepackage{natbib}
\usepackage{url}
\usepackage{placeins}
\usepackage{caption}
\usepackage{subcaption}
\def\aap{{A\&A}}

\def\aj{{AJ}}

\def\apj{{ApJ}}
\def\apjl{{ApJL}}

\def\ApJSupp{{ApJS}}

\def\mnras{{MNRAS}}

\def\apjs{{ApJS}}

\def\aaps{{A\&AS}}

\newcommand{\eg}{{e.g., }}
\newcommand{\ie}{{i.e., }}

\usepackage[normalem]{ulem}
\definecolor{gray}{rgb}{0.5,0.5,0.5}

\title[Star-galaxy Classification Using ConvNets]{Star-galaxy Classification Using Deep Convolutional Neural Networks}

\author[E. J. Kim and R. J. Brunner]{
  Edward J.~Kim$^1$\thanks{jkim575@illinois.edu} and Robert J.~Brunner$^{1,2,3,4}$ \\
$^1$Department of Physics, University of Illinois, Urbana, IL 61801 USA\\
$^2$Department of Astronomy, University of Illinois, Urbana, IL 61801 USA\\
$^3$Department of Statistics, University of Illinois, Champaign, IL 61820 USA\\
$^4$National Center for Supercomputing Applications, Urbana, IL 61801 USA }

\date{Accepted XXX. Received YYY; in original form ZZZ}

\pubyear{2016}

\begin{document}
\label{firstpage}
\pagerange{\pageref{firstpage}--\pageref{lastpage}}
\maketitle

\begin{abstract}
Most existing star-galaxy classifiers use the reduced summary information from catalogs,
requiring careful feature extraction and selection.
The latest advances in machine learning that use deep convolutional neural networks
allow a machine to automatically learn the features directly from data,
minimizing the need for input from human experts.
We present a star-galaxy classification framework that uses
deep convolutional neural networks (ConvNets) directly on the reduced, calibrated pixel values.
Using data from the Sloan Digital Sky Survey (SDSS) and
the Canada-France-Hawaii Telescope Lensing Survey (CFHTLenS),
we demonstrate that ConvNets are able to produce accurate and well-calibrated
probabilistic classifications that are competitive with
conventional machine learning techniques.
Future advances in deep learning may bring more success with current and
forthcoming photometric surveys, such as the Dark Energy Survey (DES) and
the Large Synoptic Survey Telescope (LSST), because deep neural networks require
very little, manual feature engineering.
\end{abstract}


\section{Introduction}
  \label{sec:introduction}

Currently ongoing and forthcoming large-scale photometric surveys,
such as the Dark Energy Survey (DES) and the Large Synoptic Survey Telescope (LSST),
aim to collect photometric data for hundreds of millions to billions of
stars and galaxies.
Due to the sheer volume of data, it is not possible for human experts to
manually classify them,
and the separation of photometric catalogs into stars and galaxies has to be automated.
Furthermore, any classification approach must be probabilistic in nature.
A fully probabilistic classifier enables a user to adopt probability cuts
to obtain a pure sample for population studies,
or to optimize the allocation of observing time by selecting objects for follow-up.
Ideally, however, the probability estimates themselves would be retained for all sources
and used in subsequent analyses to improve or enhance a particular
measurement~\citep{ross2011ameliorating,seo2012acoustic}.

With machine learning, we can use algorithms to automatically create
accurate source catalogs with well-calibrated posterior probabilities.
Machine learning techniques have been a popular tool
in many areas of astronomy~\citep{ball2008robust,
banerji2010galaxy,
carrascokind2013tpz,
ivezic2014statistics,
kamdar2016machine1}. 
Artificial neural networks were first applied to the problem of star-galaxy
classification in the work of \citet{odewahn1992automated},
and they have become a core part of the astronomical image processing software
\texttt{SExtractor}~\citep{bertin1996sextractor}.
Other successfully implemented examples of applying machine learning to
the star-galaxy classification problem include decision 
trees~\citep{weir1995automated,suchkov2005census,ball2006robust,vasconcellos2011decision,sevilla2015effect}, 
Support Vector Machines~\citep{Fadely2012}, and
classifier combination strategies~\citep*{kim2015hybrid}.

Almost all star-galaxy classifiers published in the literature use
the reduced summary information available from astronomical catalogs.
Constructing catalogs requires careful engineering and considerable domain expertise
to transform the reduced, calibrated pixel values that comprise an image into suitable features,
such as magnitudes or shape information of an object.
In a branch of machine learning called \textit{deep learning}~\citep{lecun2015deep},
features are not designed by human experts;
they are learned directly from data by deep neural networks.
Deep learning methods learn multiple levels of features by
transforming the feature at one level into a more abstract feature at a higher
level.
For example, when an array of pixel values is used as input to a deep learning
method, the features in the first layer might represent locations and orientations
of edges.
The second layer could assemble particular arrangements of edges into more
complex shapes, and subsequent layers would detect objects as
combinations of low-level features.
These multiple layers of abstraction progressively amplify aspects of the input
that are important for classification tasks.
Deep learning has been applied successfully to galaxy morphological
classification in Sloan Digital Sky Survey~\citep[SDSS;][]{dieleman2015rotation}
and Cosmic Assembly Near-infrared Deep Extragalactic Legacy
Survey~\citep[CANDELS;][]{huertas2015catalog} and to photometric redshift
estimation~\citep{hoyle2015measuring},
but it has not yet been applied to the problem of source classification.

In this paper, we present a star-galaxy classification framework
that uses a convolutional neural network (ConvNet) model directly on the images
from the SDSS and the Canada-France-Hawaii Telescope Lensing Survey (CFHTLenS).
We compare its performance with a standard machine learning technique
that uses the reduced summary information from catalogs,
and we demonstrate that our ConvNet model is able to produce accurate and
well-calibrated probabilistic classifications with very little feature
engineering by hand.
In Section~\ref{sec:data}, we describe the data sets used in this paper
and the pre-processing steps for preparing the image data sets.
We provide a brief overview of deep learning and ConvNets
in Section~\ref{sec:deep_learning}, and
discuss our strategy for preventing overfitting in Section~\ref{sec:overfitting}.
In Section~\ref{sec:tpc}, we describe a state-of-the-art tree-based
machine learning algorithm, to which the performance of our ConvNet model is
compared.
We present the main results of our ConvNet model in
Section~\ref{sec:results_and_discussion}, and
we outline our conclusions in Section~\ref{sec:conclusions}.

\section{Data}
  \label{sec:data}

To demonstrate the performance of our ConvNet model, we use photometric and
spectroscopic data sets with different characteristics and compositions.
In this section, we briefly describe these data sets and the image pre-processing
steps for retrieving cutout images.

\subsection{Sloan Digital Sky Survey}
  \label{sec:sdss}

The Sloan Digital Sky Survey~\citep[SDSS;][]{york2000sloan}
phases I--III obtained photometric data in five bands,
$u$, $g$, $r$, $i$, and $z$,
covering 14,555 square degrees, more than one-third of the entire sky.
The resulting catalog contains photometry of over 300 million stars and galaxies
with a limiting magnitude of $r \approx 22$,
making the SDSS one of the largest sky surveys ever undertaken.
The SDSS also conducted an expansive spectroscopic follow-up of
more than three million stars and galaxies~\citep{eisenstein2011sdss}.
In this paper, we use a subset of the photometric and spectroscopic data
contained within the Data Release 12~\citep[DR12;][]{alam2015eleventh},
which is publicly available through the online CasJobs
server\footnote{http://skyserver.sdss.org/casjobs/}~\citep{li2008casjobs}.

Using the CasJobs server, we randomly select a total of 65,000 sources,
which are either stars or galaxies.
In this work, we exclude objects that clearly are neither stars nor galaxies.
Most of the excluded objects are QSOs or quasars.
Quasars appear as point sources, rather than resolved sources similar to
galaxies,
and many of them have one or more saturated pixels in the images.
However, unlike any known stars, their spectra show strong and broad emission
lines.
Quasars are also different from galaxies because of their intrinsic variability
on a wide range of time scales, which may be due to variation in the accretion
rate or instabilities of the accretion disk around the black hole
\citep{popovic2012photocentric}.
Thus, many studies exclude quasars in the binary star-galaxy classification
scheme~\citep[\eg][]{vasconcellos2011decision,Fadely2012}.
Expanding the historical star-galaxy classification problem to include
additional classes, \eg \textit{nsng} (neither star nor galaxy),
may have advantages~\citep{ball2006robust}, and we plan to present
the results of this multi-class problem in a future paper.

We also exclude some bad photometric observations as follows.
We consider only objects
with no warning flags in the spectroscopic measurement (\texttt{zWarning = 0});
the half-light radius in the $r$ band is less than 30 arc seconds
as measured by the exponential and de Vaucouleurs light profiles;
the error on the spectroscopic redshift measurement is less than 0.1; and
the spectroscopic redshift is less than 2.

To create training images,
we obtain the image FITS files for SDSS fields containing these objects
in five photometric bands: $u$, $g$, $r$, $i$, and $z$.
We use the astrometry information in the FITS headers
in the \texttt{Montage}\footnote{http://montage.ipac.caltech.edu/} software
to align each image to the reference ($r$-band) image.
We then use \texttt{SExtractor} to find the pixel positions of
the 65,000 objects we have selected,
and to center each object in the cutout image.
Magnitudes in the SDSS photometric catalog are expressed as inverse hyperbolic
sine magnitudes~\citep[also known as luptitudes;][]{lupton1999modified},
and we follow the SDSS convention and convert all flux values to luptitudes.
Finally, in order to account for the effect of Galactic dust,
extinction corrections in magnitudes are applied
following \cite{schlegel1998maps}.
In the end, we have cutout images of size $48\times48$ pixels
with luptitude values in each pixel.
We note that we have experimented with increasing the pixel dimensions to
$60\times60$ and $72\times72$ pixels, but do not find noticeable
improvement in the performance of our model.

In the end, we have 17,344 stars and 47,656 galaxies available
for the training and testing processes.
The apparent magnitudes range from $10.7 < r < 23.1$,
and the galaxies in this sample have a mean redshift of $z \sim 0.36$.
We randomly split the objects into training, held-out validation,
and blind test sets of size 40,000, 10,000, and 15,000, respectively.
We note that cross-validation is often avoided in deep learning
in favor of hold-out validation,
since cross-validation is computationally expensive.
We also note that we perform a blind test, and the test set is not used in any
way to train or calibrate the algorithms.
The first two panels of Figure~\ref{fig:sdss_mag} show the number of objects
and the fraction of stars in the test set as functions of $r$-band magnitude.
Similarly, Figure~\ref{fig:sdss_g_r} shows the number of objects and the
fraction of stars in the test set as functions of $g-r$ color.
The normalized kernel density estimate distributions for the training
and validation sets are almost identical to those of the test set,
and they are nearly indistinguishable when overlapped.
We do not show the distributions for the training and validation sets in
Figures~\ref{fig:sdss_mag} and \ref{fig:sdss_g_r} to avoid cluttering the
plots.

\subsection{Canada-France-Hawaii Telescope Lensing Survey}

We also use photometric data from
the Canada-France-Hawaii Telescope Lensing Survey
\cite[CFHTLenS\footnote{http://www.cfhtlens.org/};][]
{heymans2012cfhtlens,erben2013cfhtlens,hildebrandt2012cfhtlens}.
This catalog consists of more than twenty five million objects
with a limiting magnitude of $i_{\text{AB}} \approx 25.5$. 
It covers a total of 154 square degrees
in the four fields (named W1, W2, W3, and W4)
of the CFHT Legacy Survey~\citep[CFHTLS;][]{gwyn2012canada}
observed in the five photometric bands:
$u$, $g$, $r$, $i$, and $z$.

We have cross-matched reliable spectroscopic galaxies from
the Deep Extragalactic Evolutionary Probe Phase 2~
\citep[DEEP2;][]{davis2003science,newman2013deep2},
the Sloan Digital Sky Survey Data Release 10~\citep[SDSS-DR10]{alam2015eleventh},
the VIsible imaging Multi-Object Spectrograph (VIMOS)
Very Large Telescope (VLT) Deep Survey~
\citep[VVDS;][]{le2005vimos,garilli2008vimos}, and
the VIMOS Public Extragalactic Redshift
Survey~\citep[VIPERS;][]{garilli2014vimos}.
We have selected only sources with very secure
redshifts and no bad flags (quality flags -1, 3, and 4 for DEEP2;
quality flag 0 for SDSS; quality flags 3, 4, 23, and 24 for VIPERS
and VVDS).

We obtain FITS images for each 1 square degree CFHTLenS pointing
that contains objects with spectroscopic labels.
We create cutout images of size $96\times96$ pixels by using a similar method
to that described in Section~\ref{sec:sdss}.
Finally, images are downscaled to $48\times48$ pixels
to reduce the computational cost.

In the end, we have 8,545 stars and 57,843 galaxies available
for the training and testing processes.
The apparent magnitudes range from $13.9 < r < 25.6$,
and the galaxies in this sample have a mean redshift of $z \sim 0.59$.
We randomly split the objects into training, held-out validation, and
blind test sets of size 40,000, 10,000, and 13,278, respectively.
Figures~\ref{fig:clens_mag} and \ref{fig:clens_g_r} show the distribution
of objects in the test set as functions of $i$-band magnitude and $g-r$ color.
We do not show the distributions for the training and validation sets,
since the normalized kernel density estimate distributions for the training
and validation sets are almost identical to those of the test set.

\section{Deep Learning}
  \label{sec:deep_learning}

Neural networks have many hyperparameters, including those that specify the
network itself (\eg the size and non-linearity of each layer)
and those that specify how the network is trained
(\eg the mini-batch size or the learning rate).
Furthermore, the architecture of a neural network can have a
significant impact on its performance.
In this section, we provide a brief description of key hyperparameters in our
ConvNet model, and also present the network architecture.

\subsection{Neural Networks}

\begin{figure*}
  \centering
  \begin{subfigure}[]{0.49\linewidth}
    \centering
    \includegraphics[width=0.6\textwidth]{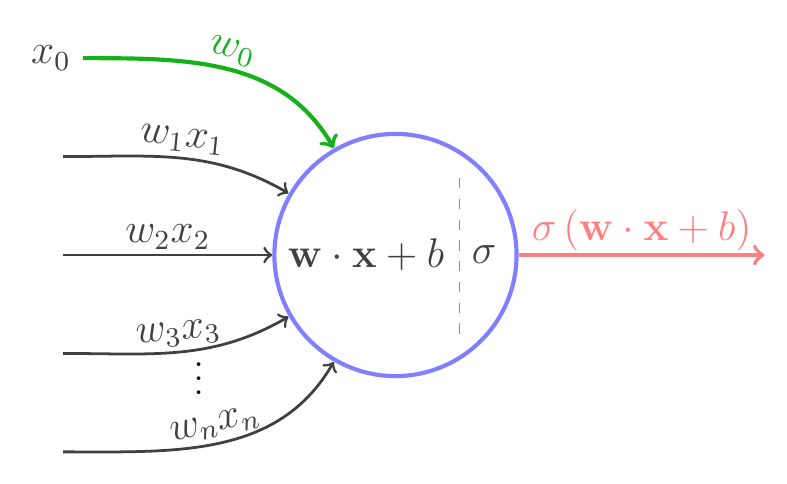}
    \caption{}
    \label{fig:neuron_a}
  \end{subfigure}
  \begin{subfigure}[]{0.49\linewidth}
    \centering
    \includegraphics[width=0.6\textwidth]{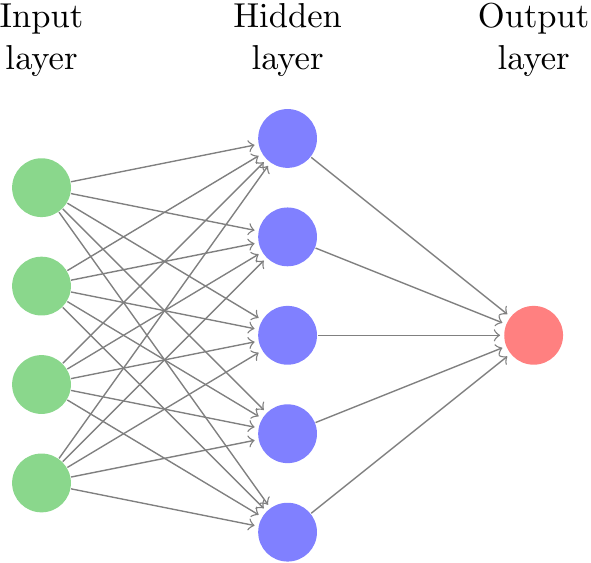}
    \caption{}
    \label{fig:neuron_b}
  \end{subfigure}
  \caption{
    (a) A mathematical model of a biological neuron.
    (b) A schematic diagram of a neural network with one hidden layer.
    }
\end{figure*}

An artificial neuron in most artificial neural networks is represented
as a mathematical function that models a biological neural structure
\citep{aggarwal2014data}.
A schematic representation is shown in Figure~\ref{fig:neuron_a}.
Let $\bmath{x}=\left(x_1,x_2,\dots,x_n\right)$ be a vector of inputs to a given neuron,
$\bmath{w}=\left(w_1,w_2,\dots,w_n\right)$ be a vector of weights, and
$b$ be the bias.
Then, the output of the neuron is
\citep{rosenblatt1961principles}
\begin{equation}
  y = \sigma \left( \bmath{w} \cdot \bmath{x} + b \right),
  \label{eq:neuron_output}
\end{equation}
where $\sigma$ is the activation function (or \textit{non-linearity}).
The most popular non-linearity at present
is the rectified linear unit
\citep[ReLU;][]{nair2010rectified}, $\sigma(x)=\max(0, x)$.
ReLUs generally allow much faster
training of deep neural networks with many layers.
However, ReLU units can sometimes result in dead neurons whose output is always zero.
To mitigate this problem, we use leaky ReLUs
~\citep{maas2013rectifier}
that have a small, non-zero slope in the negative region,
\begin{equation}
  \sigma(x) =
    \begin{cases}
      x & \mbox{if } x \geq 0 \\
      0.01x & \mbox{if } x < 0.
    \end{cases}
\end{equation}

Many deep learning models use feedforward neural network architectures with multiple layers,
where each neuron in one layer is connected to the neurons of the subsequent
layer~\citep{lecun2015deep}.
A schematic representation is shown in Figure~\ref{fig:neuron_b}.
All layers except the input and output layers are conveniently called hidden layers.

We
find a set of weights and biases such that,
given $N$ samples, the output from the network
$\bmath{y}=\left(y_1, y_2, \dots, y_N \right)$ approximates the desired output
$\hat{\bmath{y}}=\left(\hat{y}_1, \hat{y}_2, \dots, \hat{y}_N \right)$
as closely as possible for all input
$\bmath{X}=\left(\bmath{x}_1,\bmath{x}_2,\dots,\bmath{x}_N\right)$.
We can formulate this as the minimization of a loss function
$L(\bmath{y},\hat{\bmath{y}})$
over the training data.
In this work, we use \textit{cross-entropy}
\citep[also called log loss;][]{murphy2012machine} as the loss function.
For binary classification, the cross-entropy per sample is given by
\begin{equation}
  L(y_j, \hat{y}_j) = -\hat{y}_j \log_2 y_j - (1 - \hat{y}_j) \log_2(1 - y_j),
  \label{eq:cross_entropy}
\end{equation}
where $\hat{y}_j$ is the actual truth value (\eg 0 or 1) of the $j$-th data, and
$y_j$ is the probability prediction made by the model.
We compute the loss function by taking the average of all cross-entropies in the sample.
Thus, the loss function becomes
\begin{equation}
  L(\bmath{y}, \hat{\bmath{y}})
= - \frac{1}{N} \sum_{j=1}^{N} \hat{y}_j  \log_2 y_j
    + (1 - \hat{y}_j)  \log_2 (1 - \hat y_j).
  \label{eq:cross_entropy_all}
\end{equation}

To find the weights $\bmath{w}$ and biases $\bmath{b}$ which minimize the loss,
we use a technique called \textit{gradient descent},
where we use the following rules to update the parameters in each layer $l$:
\begin{align}
  \bmath{w}_l &\rightarrow
  \bmath{w}_l^{\prime}
  = \bmath{w}_l - \eta \frac{\partial L}{\partial \bmath{w}_l} \nonumber \\
  \bmath{b}_l &\rightarrow
  \bmath{b}_l^{\prime}
  = \bmath{b}_l - \eta \frac{\partial L}{\partial \bmath{b}_l},
  \label{eq:gradient_descent}
\end{align}
where $\eta$ is a small, positive number known as the \textit{learning rate}.
The gradients can be computed using the backpropagation
procedure~\citep{rumelhart1988learning}.
A common approach to speed up training is to split the training data
into mini-batches~\citep{lecun1998efficient}.
In mini-batch gradient descent, instead of computing the gradients in
Equation~\ref{eq:gradient_descent} for the entire training data,
we only compute the gradient of randomly chosen training examples at each step.
As training examples are usually correlated,
the gradient computed from each mini-batch is a good approximation
of the overall gradient~\citep{bottou1998online}.
As a result, mini-batch gradient descent results in much faster convergence.
However, there is a trade-off: the lower the batch size is,
the lower the convergence rate will be;
the higher the batch size is, the longer it will take to compute the gradient
at each step~\citep{bousquet2008tradeoffs}.
Thus, a moderate batch size, combined with a decaying learning rate, is
generally used in practice. We use a batch size of 128 in this work.

We define an \textit{epoch} as a single, complete pass through the training data,
and full training usually requires many epochs.
At the end of each epoch, we evaluate the loss function on the validation set,
and the model that minimizes the validation loss is chosen as the best model.

\subsection{Convolutional Neural Networks}
  \label{sec:convnet}

The convolutional neural network~\citep[ConvNet;][]{fukushima1980neocognitron,lecun1998gradient}
is a type of deep, feedforward neural network
that has recently become
a popular
approach in the computer vision community.
In a typical ConvNet, the first few stages are composed of two types of layers: 
convolutional layers and pooling layers.

The input to a convolutional layer is an image, and the output channels of
each layer are called \textit{feature maps}.
To produce output feature maps, we convolve each feature map with a set of weights
called \textit{filters},
and apply a non-linearity such as ReLU to the weighted sum of these convolutions.
Different feature maps use different sets of filters,
but all neurons in a feature map share the same set of filters.
Mathematically, we replace the dot product in Equation~\ref{eq:neuron_output}
with a sum of convolutions. Thus, the $k$-th feature map is given by
\begin{equation}
  y^k = \sigma \left( \sum_{m} \bmath{w}_m^k \ast \bmath{x}_m + b^k \right),
\end{equation}
where we sum over the set of input feature maps,
$\ast$ is the convolution operator, and $\bmath{w}_m^k$ represent the filters.

Typically, a
pooling layer computes the maximum of a local $2\times2$ patch in a feature map
\citep{krizhevsky2012imagenet}.
Since the pooling layer aggregates the activations of neighboring units from the previous layer,
it reduces the dimensionality of the feature maps and
makes the model invariant to small shifts and distortions
\citep{boureau2010theoretical}
.
Two or more layers of convolution and pooling are stacked,
followed by more convolutional and fully-connected layers.

\subsection{Neural Network Architecture}

\begin{table*}
  \centering
  \caption{Summary of ConvNet architecture and hyperparameters. Note that pooling layers have no learnable parameters.}
  \label{table:hyperparamters}
  \begin{tabular}{ccccccc}
    \hline
    type            & filters & filter size & padding & non-linearity & initial weights & initial biases \\
    \hline
    convolutional   & 32          & $5\times5$  & -       & leaky ReLU    & orthogonal      & 0.1            \\
    convolutional   & 32          & $3\times3$  & 1       & leaky ReLU    & orthogonal      & 0.1            \\
    pooling         & -           & $2\times2$  & -       & -             & -               & -              \\
    convolutional   & 64          & $3\times3$  & 1       & leaky ReLU    & orthogonal      & 0.1            \\
    convolutional   & 64          & $3\times3$  & 1       & leaky ReLU    & orthogonal      & 0.1            \\
    convolutional   & 64          & $3\times3$  & 1       & leaky ReLU    & orthogonal      & 0.1            \\
    pooling         & -           & $2\times2$  & -       & -             & -               & -              \\
    convolutional   & 128         & $3\times3$  & 1       & leaky ReLU    & orthogonal      & 0.1            \\
    convolutional   & 128         & $3\times3$  & 1       & leaky ReLU    & orthogonal      & 0.1            \\
    convolutional   & 128         & $3\times3$  & 1       & leaky ReLU    & orthogonal      & 0.1            \\
    pooling         & -           & $2\times2$  & -       & -             & -               & -              \\
    fully-connected & 2048        & -           & -       & leaky ReLU    & orthogonal      & 0.01            \\
    fully-connected & 2048        & -           & -       & leaky ReLU    & orthogonal      & 0.01            \\
    fully-connected & 2           & -           & -       & softmax       & orthogonal      & 0.01            \\
    \hline
  \end{tabular}
\end{table*}

We present the overall architecture of our ConvNet model in Table~\ref{table:hyperparamters}.
The network consists of eleven trainable layers.
The first convolutional layer filters the $5\times44\times44$ input image
(\ie $44\times44$ images in five bands $ugriz$) with 32 square filters of size $5\times5\times5$.
We have also experimented with using only three bands $gri$ (for three channels of RGB) and 
four bands $ugri$ and $griz$ (corresponding to RGBA),
and using only colors, \eg $u-g$, $g-r$, $r-i$, and/or $i-z$,
but we find that using magnitudes in all five bands $ugriz$ yields the best performance.

The leaky ReLU non-linearity is applied to the output of the first convolutional layer
(and all subsequent layers), and the second convolutional layer filters it
with 32 filters of $32\times3\times3$.
In the second convolutional layer (and all subsequent convolutional layers),
we pad the input with zeros spatially on the border
(\ie the zero-padding is 1 pixel for $3\times3$ convolutional layers)
such that the spatial resolution is preserved after convolution.
Max-pooling with filters of size $2\times2$ follows the second convolutional layer.
A stack of six additional convolutional layers, all with filters of size $3\times3$,
is followed by three fully-connected layers.
The first two fully-connected layers have 2048 channels each,
and the third performs binary classification.

The output of the final fully-connected layer is fed to a \textit{softmax} function.
The softmax function is given by
\begin{equation}
  P (G \mid \bmath{x}) = 
    \frac{ e^{ \bmath{x} \cdot \bmath{w}_G } }
    { \sum_{i} e^{ \bmath{x} \cdot \bmath{w}_i } },
\end{equation}
where we sum over the different possible values of the class label (\ie star or galaxy),
and interpret its output as the posterior probability that an object is a galaxy
(or a star).
We note that we could also try to solve a regression problem, \eg
by normalizing the output values that the network produces for each class.
However, we find that solving a regression problem instead of using
the softmax function results in significantly worse performance.

We have performed a manual search to explore more than 200 combinations of different architectures
and hyperparameters to find an architecture that minimizes the loss function
(Equation~\ref{eq:cross_entropy_all}) on the validation set of the SDSS data.
The architecture described in this section provides the best performance on
the SDSS validation set.
To test how this model performs across different, related data,
we use the same architecture on the CFHTLenS data set.

The architecture of \cite{krizhevsky2012imagenet}
uses relatively large receptive fields ($11\times11$) in the first convolutional layers.
\citet{zeiler2014visualizing} and \cite{dieleman2015rotation}
also use large receptive fields of $7\times7$ and $6\times6$
in the first convolution layer, respectively.
However, we find that using a receptive field larger than $5\times5$ in the first
convolutional layer leads to worse performance.
This result is in agreement with
the network of \citet{simonyan2014very}, which has become known as ``VGGNet".
VGGNet features an extremely homogeneous architecture that only performs
$3\times3$ convolutions.
Using a large receptive field instead of a stack of multiple $3\times3$ convolutions
leads to a shallower network, and it is often preferable to increase
the depth by using smaller receptive fields.
However, we find that replacing the first layer with
a stack of two $3\times3$ convolutional layers increases the validation error,
and thus use a $5\times5$ convolution in the first layer.

In the remaining layers, we still follow VGGNet and add many
$3\times3$ convolutions (with zero-padding of size 1 pixel).
Note that with the padding of 1 pixel for $3\times3$ convolutional layers,
the spatial resolution will be preserved after convolution.
Such preservation of spatial resolution allows us to build relatively deep
networks, as shown in Table~\ref{table:hyperparamters}.
The main contribution of VGGNet is in showing that the depth plays an important
role in good performance.
In our case, we start with four convolutional layers and progressively add more
layers, while monitoring the validation loss;
we stop at eight convolutional layers after we find that adding more layers
leads to worse performance.
We conjecture that a greater depth and hence larger number of parameters
lead to overfitting in our case.

The choice of momentum, learning rate, and initial weights is crucial
for achieving high predictive performance and speeding up the learning
process~\citep{sutskever2013importance}.
To train our models, we use mini-batch gradient descent with a batch size of 128
and Nesterov momentum~\citep{bengio2013advances} of 0.9.
We initialize the learning rate $\eta$ at 0.003 for all layers and
decrease it linearly with the number of epochs 
from 0.003 to 0.0001 over 750 epochs.
We also initialize the weights in each layer with random orthogonal initial 
conditions~\citep{saxe2013exact}.
We use slightly positive values ($b=0.01$ or $0.1$) for all biases.
We find initializing biases to a small constant value helps eliminate dead neurons
by ensuring that all ReLU neurons fire in the beginning.

To implement our model, we use Python and the Lasagne
library~\citep{dieleman2015lasagne}, which is built on top of
Theano~\citep{theano2016theano}.
The Theano library simplifies the use of GPU for computation, and using the GPU
allows about an order of magnitude faster training than using just the CPU.
We note that training our network takes about forty hours on an NVIDIA Tesla
K40 GPU.
In the interest of scientific reproducibility, we make all our code available
at \mbox{\url{https://github.com/EdwardJKim/dl4astro}}.

\section{Reducing Overfitting}
  \label{sec:overfitting}
  
Our convolution neural network has $11\times10^6$ learnable parameters,
while there are only $4\times10^4$ images in the training set.
As a result, the model is very likely to \emph{overfit} without regularization.
In this section, we describe the techniques we used to minimize overfitting.

\subsection{Data Augmentation}
  \label{sec:data_augmentation}
  
One common method to combat overfitting is to artificially increase
the number of training data by using label-preserving
transformations~\citep{krizhevsky2012imagenet,dieleman2015rotation,dieleman2016exploiting}.
Each image is transformed as follows:
\begin{itemize}
\item{Rotation: 
Rotating an image does not change whether the object is a star or a galaxy.
We exploit this rotational symmetry and randomly rotate each image by a multiple of
$90^{\circ}$. }

\item{Reflection:
We flip each image horizontally with a probability of 0.5 to exploit mirror symmetry. }
\item{Translation:
We also have translational symmetry in the images.
Given an image of size $48\times48$ pixels, we extract a random contiguous crop
of size $44\times44$.
Each cropping is equivalent to randomly shifting a $44\times44$ image by up to 4 pixels
vertically and/or horizontally. }
\item{Gaussian noise:
We introduce random Gaussian noise to each pixel values
by using a similar method to \cite{krizhevsky2012imagenet}.}
\end{itemize}
In addition to artificially increasing the size of the data set,
these data augmentation schemes make the resulting model more invariant to
rotation, reflection, translation, and small noise in the pixel values.
We also note that the data augmentation steps
add almost no computational cost,
as they are performed on the CPU while the GPU is training the ConvNets on images.

\subsection{Dropout}

We use a regularization technique called dropout~\citep{hinton2012improving}
in the fully-connected layers.
Dropout consists of randomly setting to zero
the output of each hidden neuron of the previous layer with probability 0.5.
The weights of the remaining neurons are multiplied by 0.5 to preserve
the scale of input values to the next layer.
Since a neuron can be removed at any time, it cannot rely on the presence of other neurons
in the same layer and is forced to learn more robust features.

\subsection{Model Combination}
  \label{sec:bmc}

To make final classifications,
we use our ConvNet model to make 64 sets of predictions for
64 transformations of the input images:
4 rotations, 4 horizontal translations, and 4 vertical translations
(with random horizontal reflections).
Although we use an identical network architecture for all transformations,
we consider each set of predictions as separate results from different models.
Finally, we use a model combination technique known as
Bayesian Model Combination~\citep[BMC;][]{Monteith2011}, which
uses Bayesian principles to generate an ensemble combination of different models.
Although the data augmentation step in Section~\ref{sec:data_augmentation} should
make our ConvNet model invariant to these types of transformations,
we find that applying BMC still results in a significant increase in performance.
For a thorough discussion of BMC, we refer the reader to \citet{Monteith2011}
(See also \citet{carrascokind2014exhausting} for application of BMC to
photometric redshift estimation,
and \citet{kim2015hybrid} for combining star-galaxy classifiers).

\section{Trees for Probabilistic Classifications}
  \label{sec:tpc}

To compare the performance of ConvNets with
machine learning algorithms that use standard photometric features,
we use a machine learning framework called
Trees for Probabilistic Classifications (TPC).
TPC is a parallel, supervised machine learning algorithm
that uses prediction trees and random forest 
techniques~\citep{breiman1984classification, breiman2001random}
to produce a star-galaxy classification.
A complete description of TPC is beyond the scope of this paper, and
we refer the reader to
\cite{carrascokind2013tpz} and \cite{kim2015hybrid} for more details.
While other random forest implementations exist, we have chosen TPC, because
it is tailored specifically for astronomical use~\citep{carrascokind2013tpz};
it has been tested for astronomical use cases, including
photometric redshift estimation~\citep{sanchez2014photometric}
and star-galaxy classification~\citep{kim2015hybrid};
and it uses parallelism to handle large data sets on distributed memory systems.

We train two TPC models on the SDSS data set by using different sets of
attributes.
The first model, which we denote $\rmn{TPC}_{\rmn{phot}}$, is trained with
only nine photometric attributes:
the extinction-corrected model magnitudes in five bands
($u$, $g$, $r$, $i$, $z$)
and their corresponding colors
($u-g$, $g-r$, $r-i$, $i-z$).
The second model, which we denote $\rmn{TPC}_{\rmn{morph}}$, is trained with
the concentration parameter in each band in addition to the magnitudes and
colors, for a total of
fourteen
dimensions.
The concentration is defined as the difference between the
PSF magnitude (psfMag) and the composite model magnitude (cModelMag), \ie
$\rmn{concentration}\equiv\rmn{psfMag}-\rmn{cModelMag}$.
The SDSS pipeline uses a parametric method based on the concentration,
an object is classified as a galaxy if $\rmn{concentration}>0.145$.
We find that the concentration is an excellent morphological feature
for star-galaxy separation, and including more morphological features
does not show noticeable improvement in performance.
The concentration is a good example of carefully handcrafted feature extraction;
we show in Section~\ref{sec:results_and_discussion}
that ConvNets do not require such feature engineering.

We also train two models on the CFHTLenS data set.
$\rmn{TPC}_{\rmn{phot}}$ is trained with 
the five magnitudes and their corresponding colors:
$u$, $g$, $r$, $i$, $z$, $u-g$, $g-r$, $r-i$, and $i-z$.
Since the CFHTLenS catalog does not provide the concentration parameter,
$\rmn{TPC}_{\rmn{morph}}$ uses
\texttt{SExtractor}'s \texttt{FLUX\_RADIUS} (the half-light radius),
\texttt{A\_WORLD} (the semi-major axis), and \texttt{B\_WORLD}
(the semi-minor axis) for morphological features, in addition to
the five magnitudes and their corresponding colors,
for a total of twelve dimensions.

\section{Results and Discussion}
  \label{sec:results_and_discussion}

In this section, we first describe the performance metrics that were used for
evaluating the models.
We then present the classification performance of our ConvNet model
on the CFHTLenS and SDSS data sets, and compare it with the performance of TPC.

\subsection{Classification Metrics}

\begin{table}
  \caption{The definition of the classification performance metrics.}
  \centering
  \begin{tabular}{c l}
  \hline
  Metric & Meaning \\
  \hline
  AUC & Area under the Receiver Operating Curve \\
  MSE & Mean squared error \\
  $c_g$ & Galaxy completeness \\
  $p_g$ & Galaxy purity \\
  $c_s$ & Star completeness \\
  $p_s$ & Star purity \\
  $p_g(c_g=x)$ & Galaxy purity at $x$ galaxy completeness \\
  $c_s(p_s=x)$ & Star completeness at $x$ star purity \\
  $CAL$ & Calibration error with overlapping binning \\
  $\mid\Delta N_g\mid / N_g$ & Absolute error in number of galaxies\\
log loss
&
Cross-entropy
\\
  \hline
  \end{tabular}
  \label{table:metrics}
\end{table}

Probabilistic classifiers, rather than only assigning discrete labels to each
source, produce a continuous probability distribution of whether each source is
a star or a galaxy.
To evaluate the performance of probabilistic classifiers,
many studies~\citep[\eg][]{henrion2011bayesian, Fadely2012} convert
probability estimates into class labels by choosing a probability
threshold, \eg a source is classified as a star if $P_{\rmn{class}} < 0.5$,
and a galaxy if $P_{\rmn{class}} > 0.5$.
However, using a fixed threshold ignores the model's operating conditions,
such as science requirements, misclassification costs, and stellar distribution.
Furthermore, the probability threshold of 0.5 is not necessarily optimal for
an unbalanced data set, where galaxies outnumber stars.

Following \citet{kim2015hybrid}, we use performance metrics that are well-suited
for probabilistic classifiers: Area Under the Curve (AUC)
for the Receiver Operating Characteristic (ROC) curve, completeness and purity,
and the Mean Squared Error (MSE).
A good probabilistic classifier should also provide well-calibrated
posterior probabilities.
Thus, to evaluate calibration performance,
we also use the calibration error and the absolute error in the estimation
of number of galaxies.

\subsubsection{Receiver Operating Characteristic Curve}

A Receiver Operating Characteristic (ROC) curve is the
most commonly used method for evaluating the overall performance of a binary
classifier~\citep*{swets2000better}.
In an ROC curve, we plot the true positive rate as a function of the false
positive rate by varying the classification threshold.
The Area Under the Curve (AUC) quantifies the overall performance in a single
number.

\subsubsection{Completeness and Purity}

Let $N_g$ be the number of true galaxies classified as galaxies,
and $M_g$ the number of true galaxies classified as stars.
Then the galaxy \textit{completeness} $c_g$ (also called recall or sensitivity)
is given by
\begin{equation}
c_g = \frac{N_g}{N_g + M_g}.
\end{equation}
Let $M_s$ be the number of true stars classified as galaxies.
Then the galaxy \textit{purity} $p_g$ (also called precision or positive
predictive value) is given by
\begin{equation}
p_g = \frac{N_g}{N_g + M_s}.
\end{equation}
We define the star completeness and purity in a similar way.
As discussed in our previous work~\citep{kim2015hybrid},
we adopt weak lensing science requirements of the DES~\citep{soumagnac2013star},
and compute $p_g$ at $c_g=0.960$ and $c_s$ at $p_s=0.970$.

\subsubsection{Mean Squared Error}

We also use the mean squared error
(MSE; also known as the Brier score~\citep{brier1950verification})
as a performance metric. We define MSE as
\begin{equation} \label{eq:mse}
  \rmn{MSE} = \frac{1}{N} \sum^{N}_{j=1}
  \left( y_j - \hat{y}_j \right)^2,
\end{equation}
The MSE can be considered as both a score function that quantifies
how well a set of probabilistic predictions is calibrated,
or a loss function.

\subsubsection{Calibration Error}

A fully probabilistic classifier predicts not only the class label,
but also its confidence level on the prediction.
In a well-calibrated classifier, the posterior class probability estimates
should coincide with the proportion of objects that truly belong to a certain class.
Probability \textit{calibration curves}~\citep[or reliability curves;][]{degroot1983comparison}
are often used to display this relationship,
where we bin the probability estimates and plot the
fraction of positive examples versus the predicted probability in each bin
(see Figures~\ref{fig:clens_calibration} and \ref{fig:sdss_calibration}).

The problem with a binning approach is either too few or too many bins
can distort the evaluation of calibration performance.
Thus, we adopt a calibration measure based on
overlapping binning~\citep{caruana2004data}.
We order the predicted values $P_\rmn{class}$ and
put the first 1,000 elements in the first bin.
We calculate the true probability $P_\rmn{gal}$ by counting the true galaxies
in this bin.
The calibration error for this bin is $\mid P_\rmn{gal} - P_\rmn{class} \mid$.
We then repeat this for the second bin (2 to 1,001),
the third bin (3 to 1,002), and so on, and average the binned calibration errors.
Thus, the overall calibration error is given by
\begin{equation}
  \rmn{CAL} = \frac{1}{N-s} \sum_{b=1}^{N-s} \sum_{j=b}^{b+s-1}
  \bigg| P_{\rmn{class},j} - \frac{\sum_{j=b}^{b+s-1} P_{\rmn{gal},j}}{s} \bigg|,
\end{equation}
where $s=1000$ is the bin length, which is chosen approximately equal to
the number of objects in the testing set divided by the number of bins used in
the calibration curve, \ie $s \approx N / 10$.

\subsubsection{Number of galaxies}

Ideally, the probabilistic output of a classifier would be used in subsequent
scientific analyses.
For example, one can weight each object by the probability that it is a galaxy
when measuring auto-correlation functions of luminous
galaxies~\citep{ross2011ameliorating}.
In other words, given a well-calibrated classifier,
instead of counting each galaxy equally, a galaxy could be
counted as $P_{\rmn{class}}$, the probability estimate.
This should in principle remove the contamination effect of stars.
For a perfect classifier, we can count the total number of galaxies in the sample
by summing the values of $P_{\rmn{class}}$.
Thus, we measure the reliability of classifier output by the absolute error
in the estimation of number of galaxies,
\begin{equation}
  \frac{\mid \Delta N_g \mid}{N_g} =
  \frac{\big| N_g - \sum_{j=1}^{N} P_{\rmn{class},j} \big|}{N_g}.
\end{equation}

\subsection{CFHTLenS}
  \label{sec:results_cfht}

\begin{table*}
  \caption{
    A summary of the classification performance metrics
    as applied to the CFHTLenS data.
    The definition of the metrics is summarized in Table~\ref{table:metrics}.
    The bold entries highlight the best performance values within each column.
    Note that $p_{g}(c_g=0.96)$ and $c_{s}(p_s=0.97)$ require adjusting
    threshold values (\ie probability cuts), while other metrics do not.
    To obtain a galaxy completeness of $c_g=0.96$, we choose the threshold values
    0.9972, 0.9963, and 0.9927 for ConvNet, TPC${}_{\rmn{morph}}$, and 
    TPC${}_{\rmn{phot}}$, respectively;
    for star purity $p_s=0.97$, we choose 0.6990, 0.5297, and 0.8570 for
    ConvNet, TPC${}_{\rmn{morph}}$, and TPC${}_{\rmn{phot}}$, respectively.
  }
  \centering
  \begin{tabular}{l c c c c c c c}
    \hline
    classifier & AUC & MSE & $p_{g}(c_g=0.96)$ & $c_{s}(p_s=0.97)$ & CAL & $ |\Delta N_g|/N_g$ &
log loss
\\
    \hline
    ConvNet                       & \textbf{0.9948} & 0.0112          & \textbf{0.9972} & 0.8971          & \textbf{0.0197}  & \textbf{0.0029} &
\textbf{0.0441}
\\
    TPC${}_{\rmn{morph}}$         & 0.9924          & \textbf{0.0109} & 0.9963          & \textbf{0.9268} & 0.0245           & 0.0056 &
0.0809
\\
    TPC${}_{\rmn{phot}}$          & 0.9876          & 0.0189          & 0.9927          & 0.8044          & 0.0266           & 0.0101 &
0.1085
\\
    \hline
  \end{tabular}
  \label{table:clens_metrics}
\end{table*}

\begin{figure}
  \centering
  \includegraphics[width=\columnwidth]{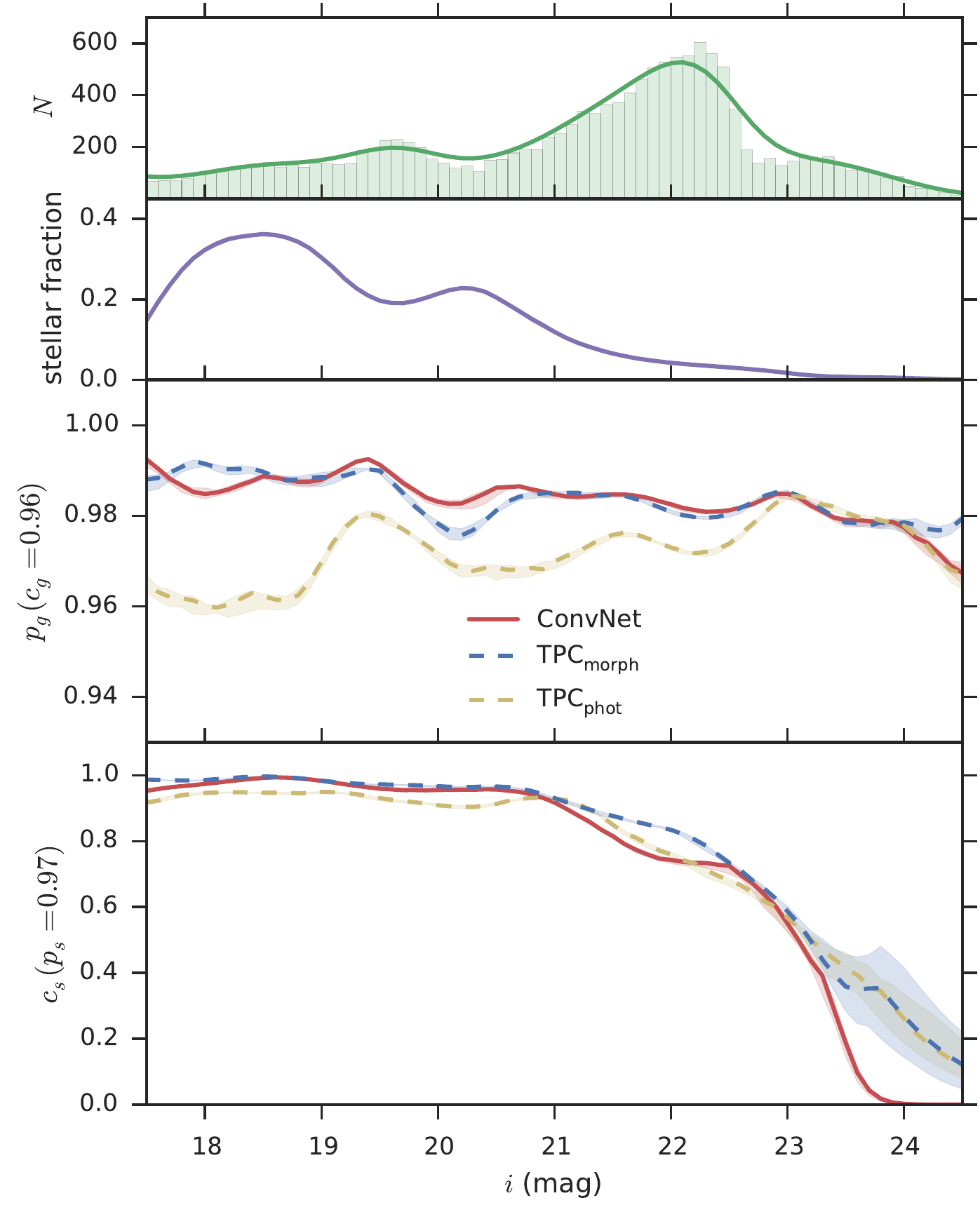}
  \caption{Galaxy purity and star completeness values as functions
           of the $i$-band magnitude (differential counts)
           as estimated by kernel density estimation (KDE)
           in the CFHTLenS data set.
           The top panel shows the histogram with a bin size of 0.1 mag
           and the KDE for objects in the test set.
           The second panel shows the fraction of stars estimated by KDE
           as a function of magnitude.
           The bottom two panels compare the galaxy purity and star completeness
           values for ConvNet (red solid line),
           $\rmn{TPC}_{\rmn{morph}}$ (blue dashed line),
           and $\rmn{TPC}_{\rmn{phot}}$ (yellow dashed line)
           as functions of magnitude.
           The $1 \sigma$ confidence bands are estimated by
           bootstrap sampling.}
  \label{fig:clens_mag}
\end{figure}

\begin{figure}
  \centering
  \includegraphics[width=\columnwidth]{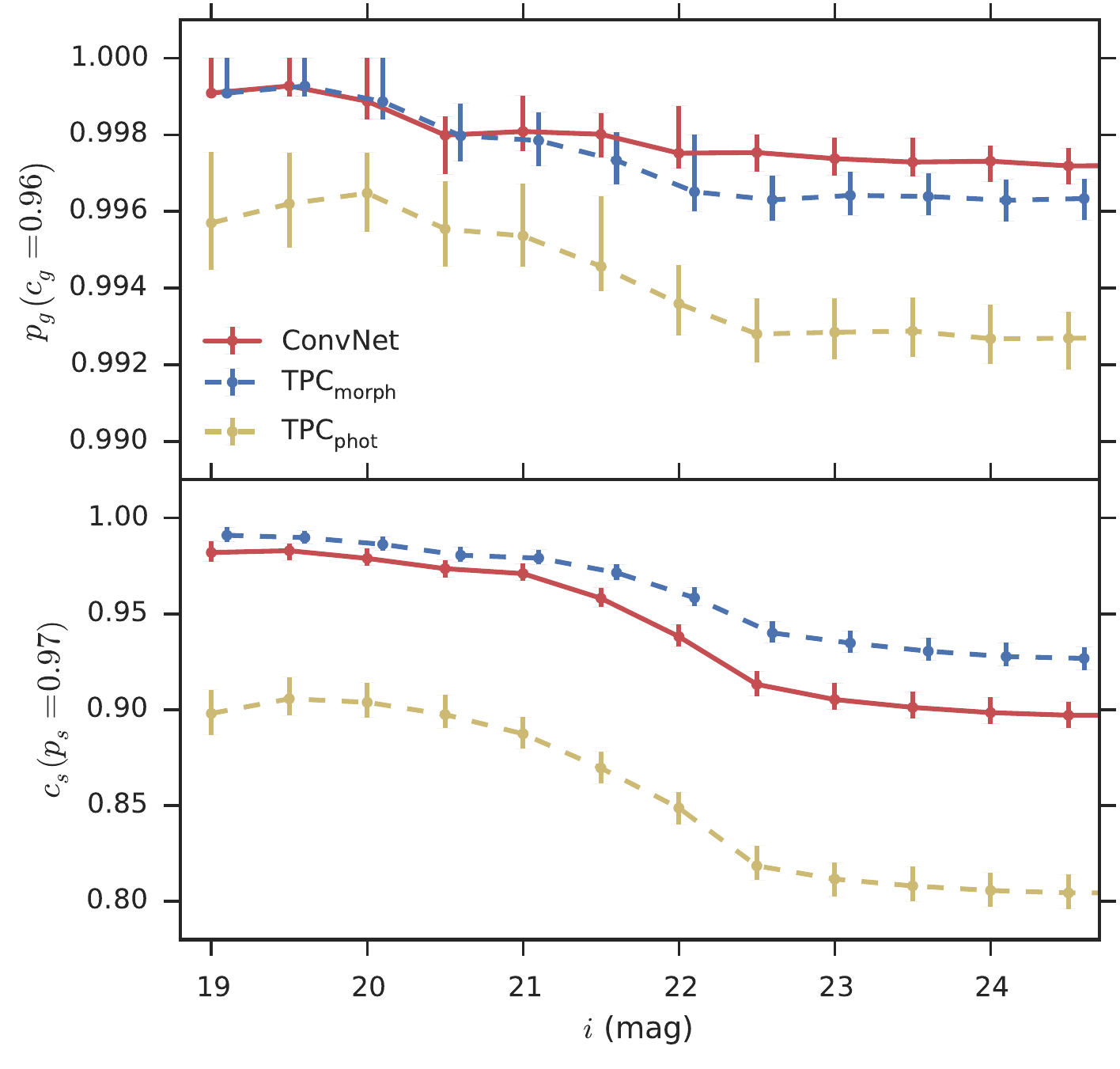}
  \caption{
    Galaxy purity and star completeness as functions of the $i$-band
    magnitude (integrated counts) in the CFHTLenS data set.
    The upper panel compares the galaxy
    purity values for ConvNet (red solid line), $\rmn{TPC}_{\rmn{morph}}$
    (blue dashed line), and $\rmn{TPC}_{\rmn{phot}}$ (yellow dashed line).
    The lower panel compares the star completeness values.
    The $1 \sigma$ error bars are computed following the method
    of \citet{paterno2004calculating} to avoid the unphysical
    errors of binomial or Poisson statistics.
    }
  \label{fig:clens_integrated}
\end{figure}

\begin{figure}
  \centering
  \includegraphics[width=\columnwidth]{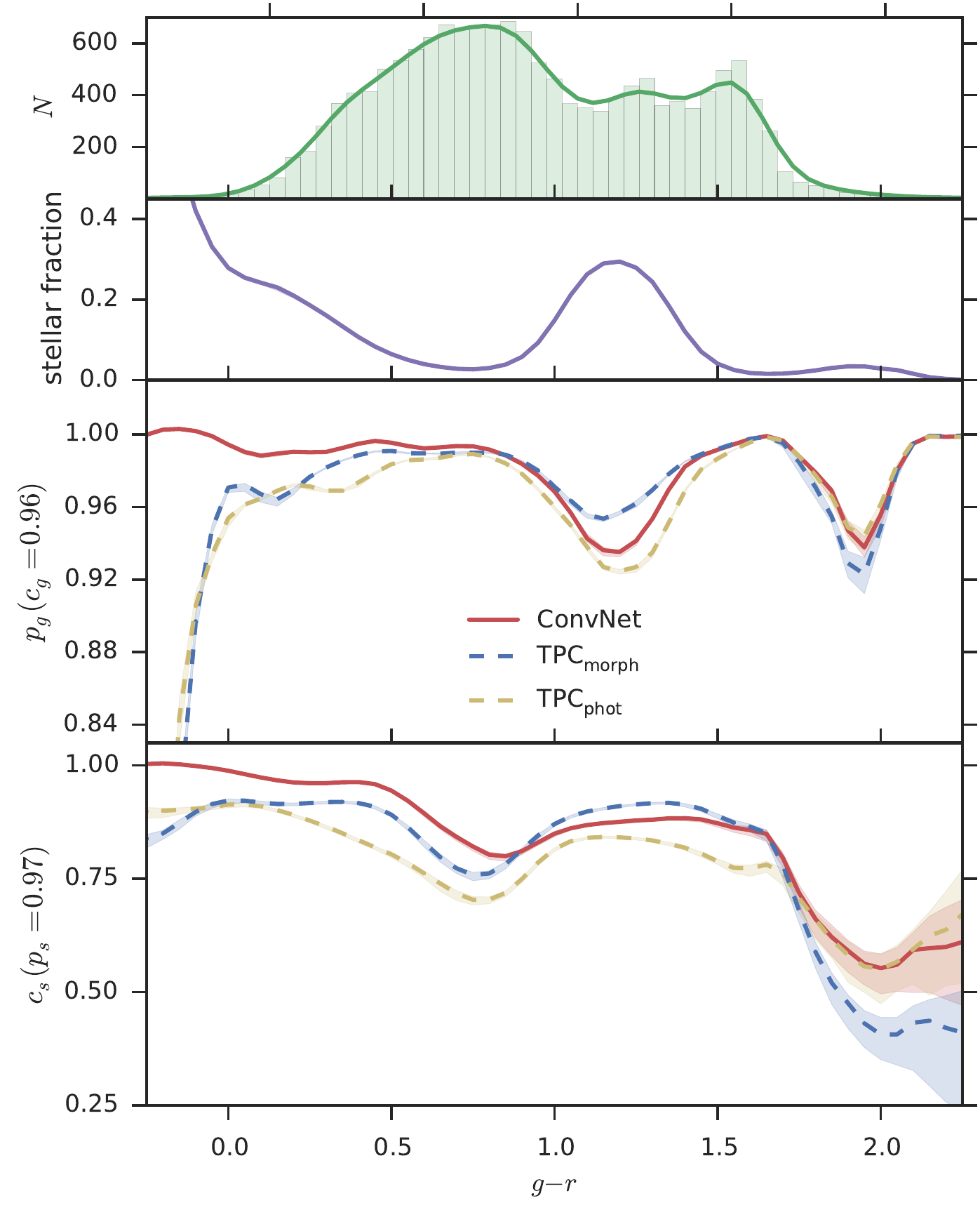}
  \caption{Similar to Figure~\ref{fig:clens_mag}
           but as a function of $g-r$ color.
           The bin size of histogram in the top panel is 0.05.}
  \label{fig:clens_g_r}
\end{figure}

\begin{figure}
  \centering
  \includegraphics[width=\columnwidth]{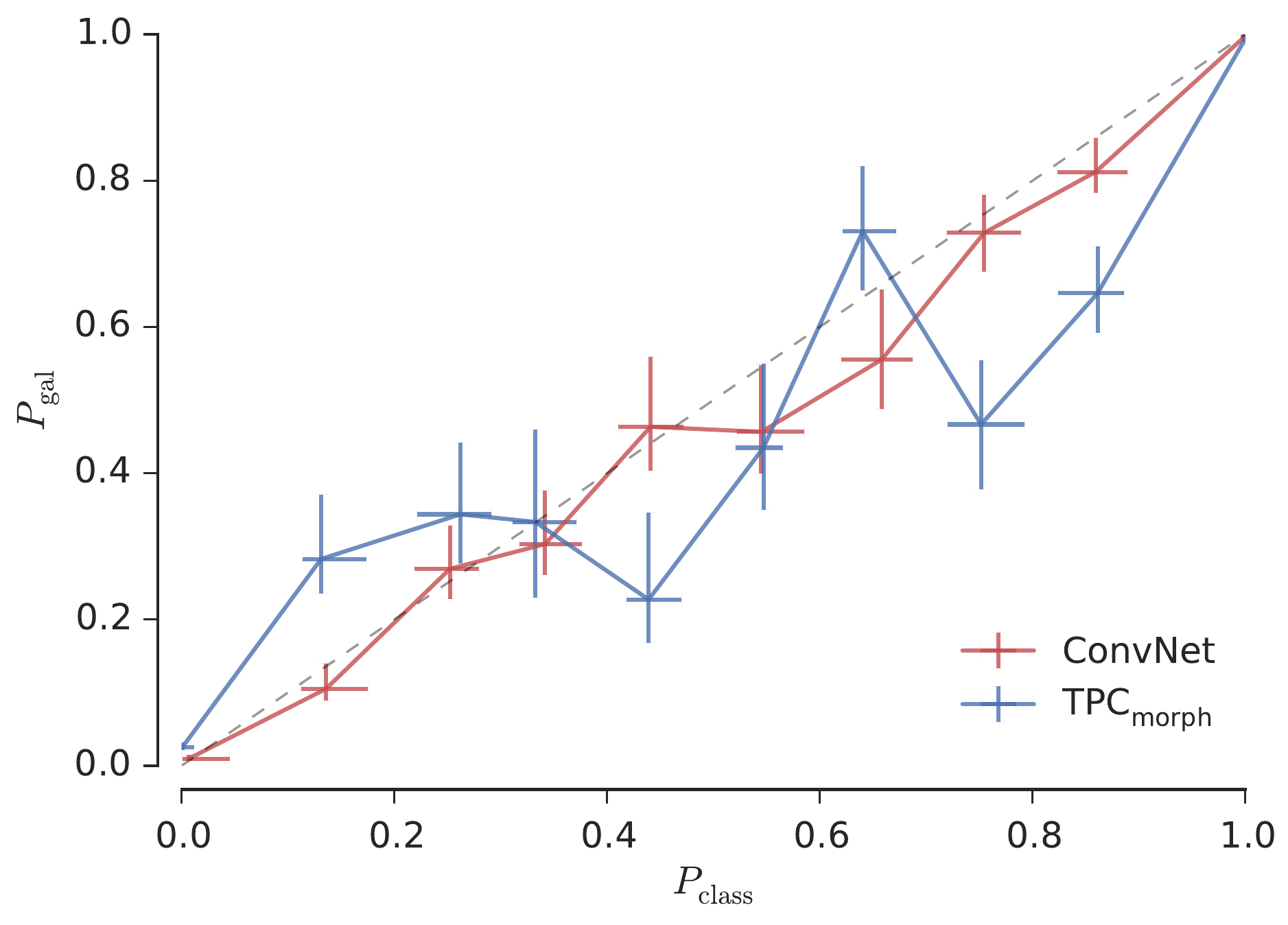}
  \caption{
    The calibration curves for ConvNet (red) and $\rmn{TPC}_{\rmn{morph}}$ (blue)
    as applied to the CFHTLenS data set.
    $P_{\rmn{gal}}$ is the fraction of objects that are galaxies, and
    $P_{\rmn{class}}$ is the probabilistic outputs generated by the classifiers.
    The dashed line displays the relationship
    $P_{\rmn{gal}} = P_{\rmn{conv}}$.
    The $1 \sigma$ error bars are computed following the method
    of \citet{paterno2004calculating}.
    }
  \label{fig:clens_calibration}
\end{figure}

\begin{figure*}
  \centering
  \begin{subfigure}[c]{0.24\linewidth}
  \centering
    \includegraphics[width=\textwidth]{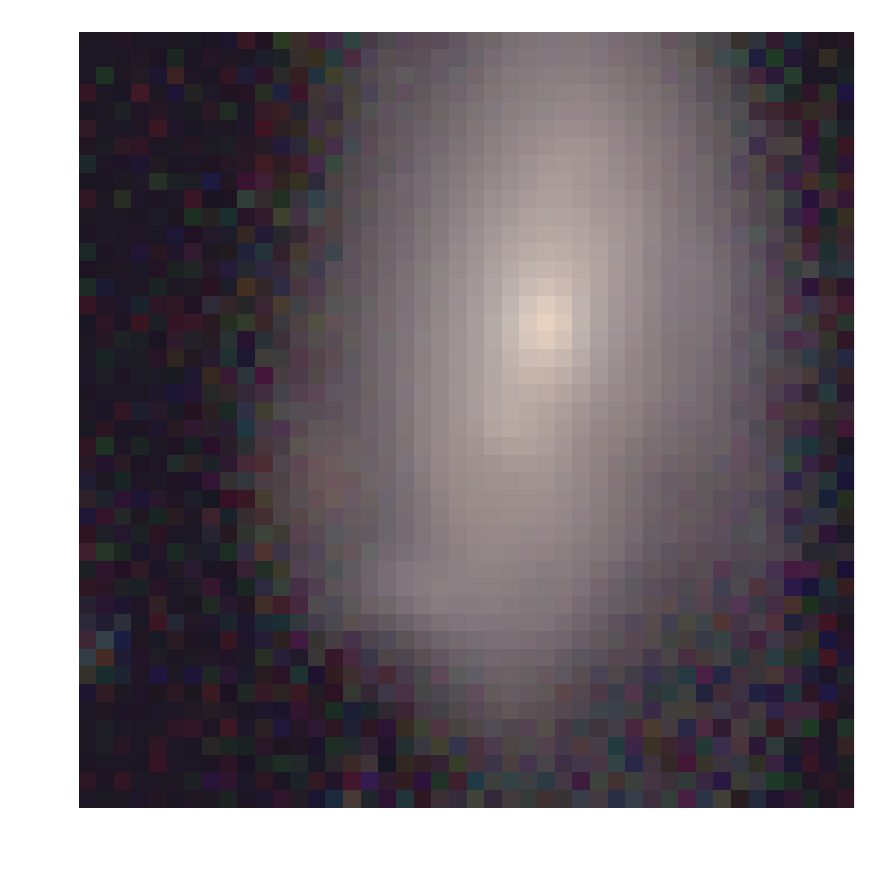}
    \caption{Input (5 bands$\times44\times44$)}
  \end{subfigure}
  \hfill
  \begin{subfigure}[c]{0.24\linewidth}
  \centering
    \includegraphics[width=\textwidth]{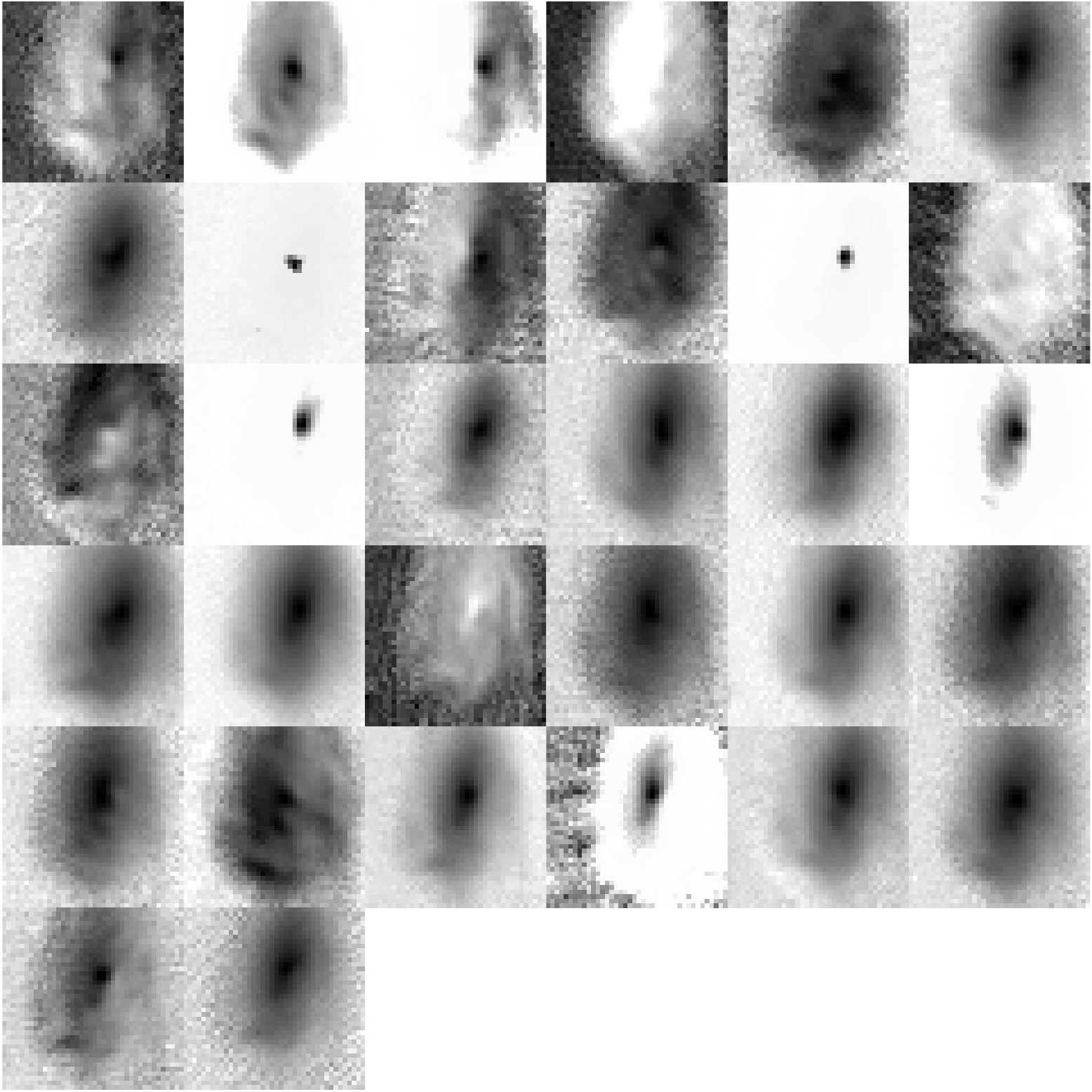}
    \caption{Layer 1 (32 maps$\times40\times40$)}
  \end{subfigure}
  \hfill
  \begin{subfigure}[c]{0.24\linewidth}
  \centering
    \includegraphics[width=\textwidth]{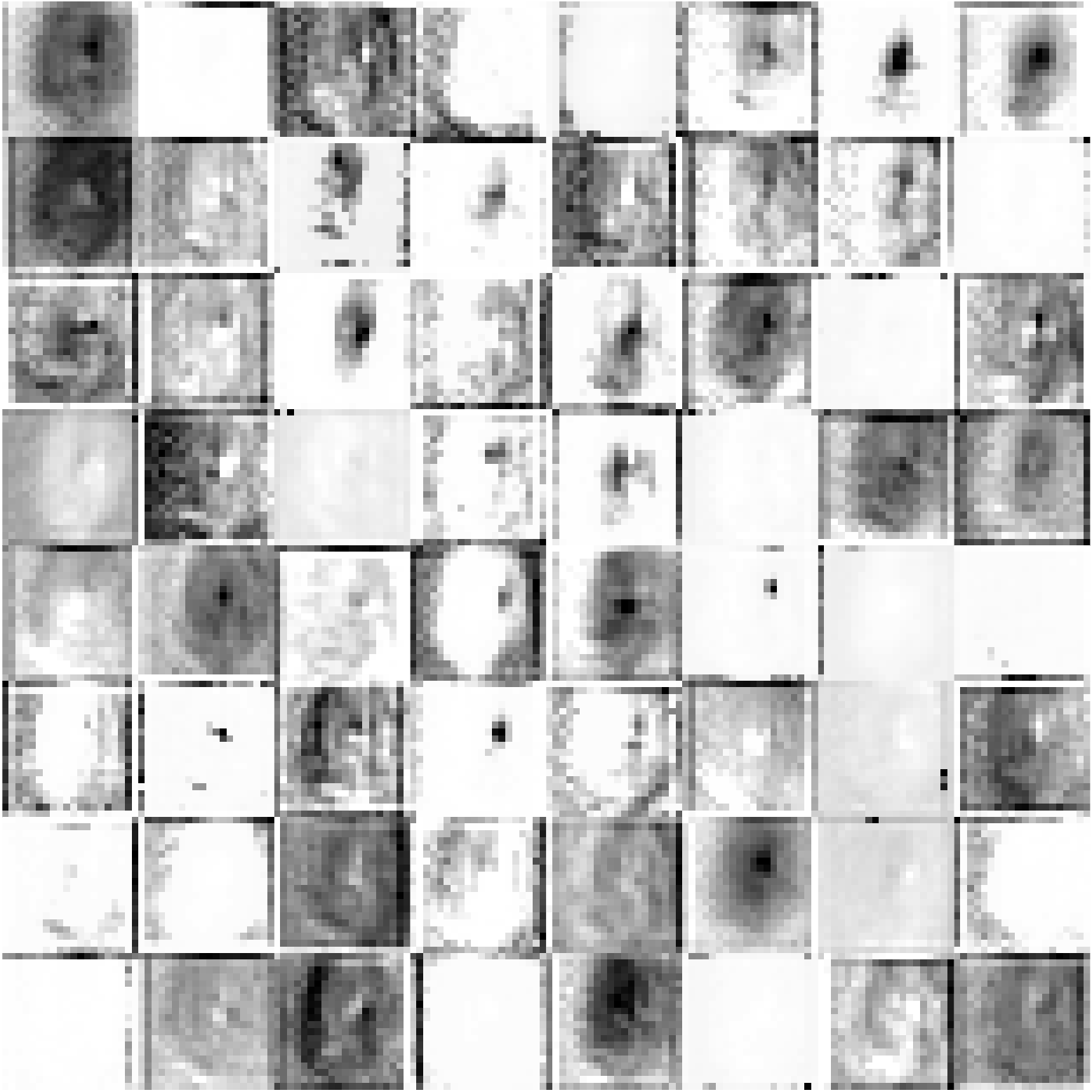}
    \caption{Layer 3 (64 maps$\times20\times20$)}
  \end{subfigure}
  \hfill
  \begin{subfigure}[c]{0.24\linewidth}
  \centering
    \includegraphics[width=\textwidth]{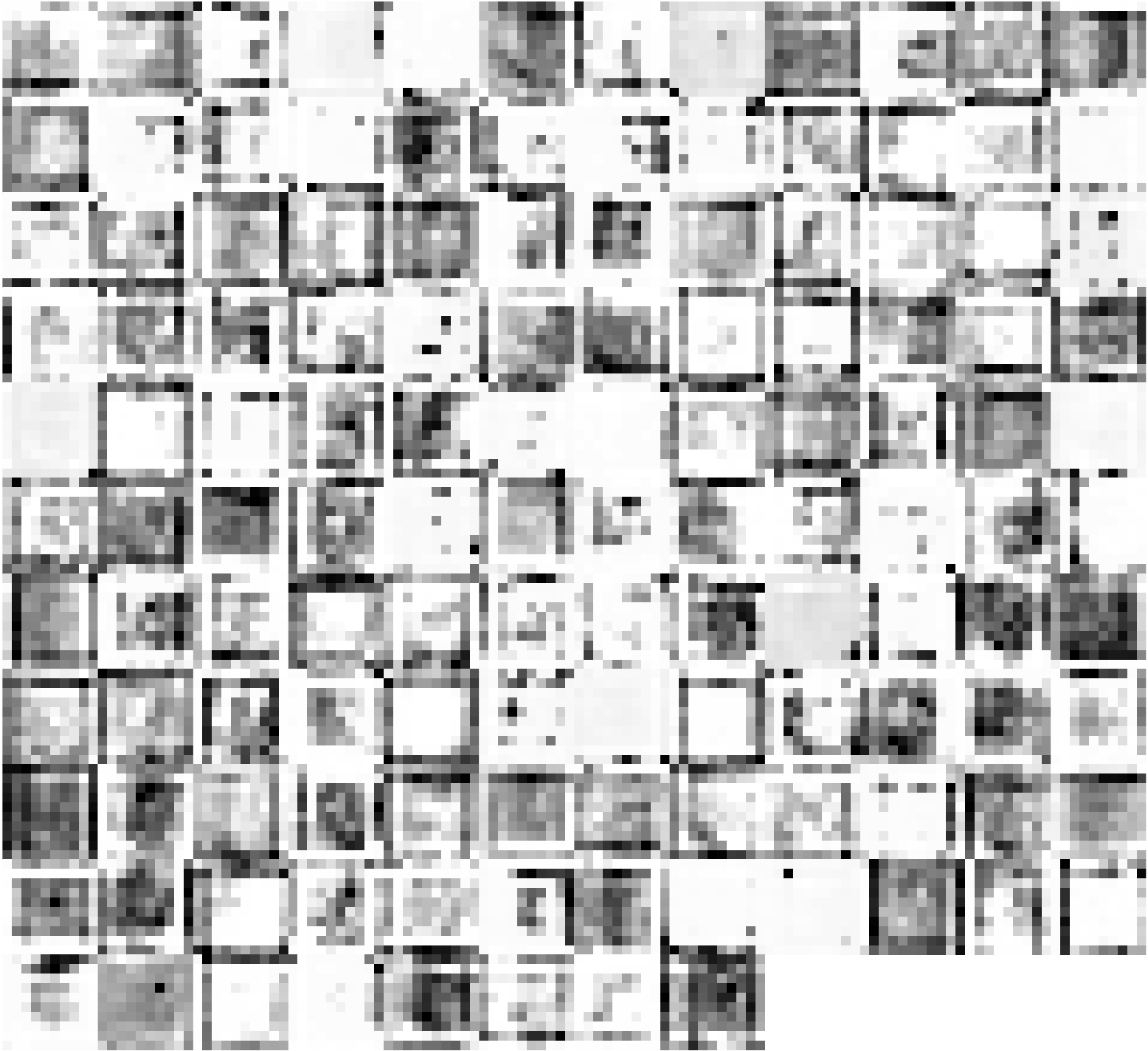}
    \caption{Layer 6 (128 maps$\times10\times10$)}
  \end{subfigure}
  \caption{
    (a) A sample $44\times44$ RGB image of a galaxy in the CFHTLenS data set.
    The RGB image is created by mapping R $\rightarrow i$ band magnitude,
    G $\rightarrow r$ band magnitude, and B $\rightarrow g$ band magnitude.
    (b) Activations on the first convolutional layer when a $5\times44\times44$
    image is fed into the network.
    (c) Activations on the third convolutional layer.
    (d) Activations on the sixth convolutional layer.
    Each image in (b), (c), and (d) is a feature map corresponding to the
    output for one of the learned features.
    }
  \label{fig:galaxy_activations}
\end{figure*}

\begin{figure*}
  \centering
  \begin{subfigure}[c]{0.24\linewidth}
  \centering
    \includegraphics[width=\textwidth]{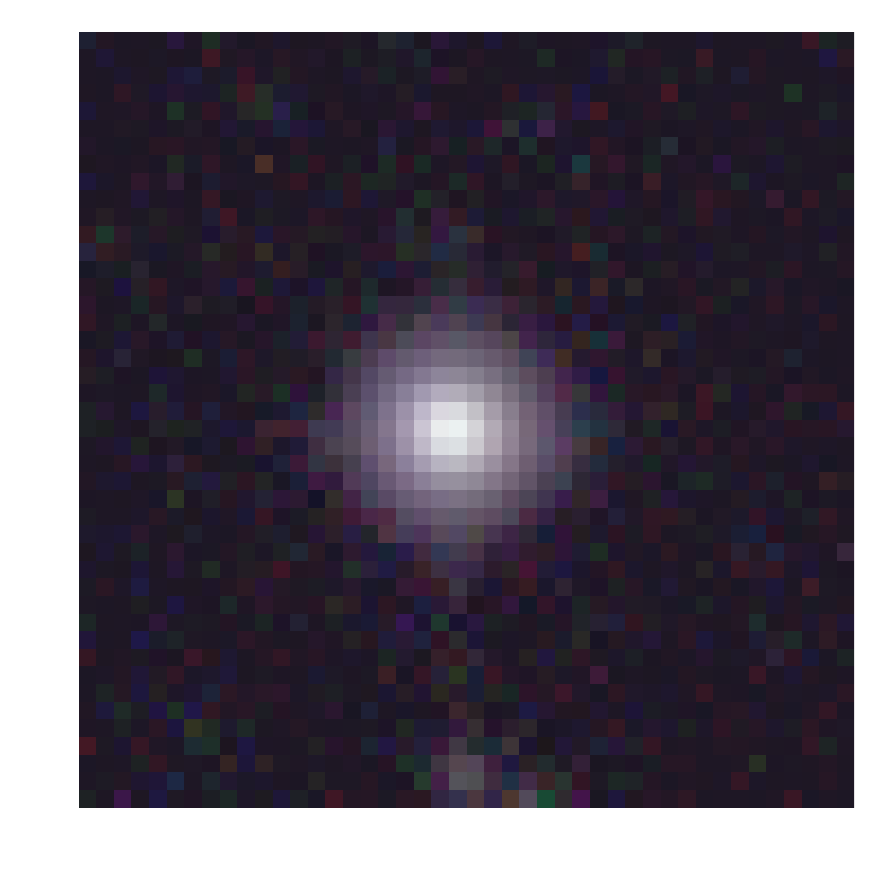}
    \caption{Input (5 bands$\times44\times44$)}
  \end{subfigure}
  \hfill
  \begin{subfigure}[c]{0.24\linewidth}
  \centering
    \includegraphics[width=\textwidth]{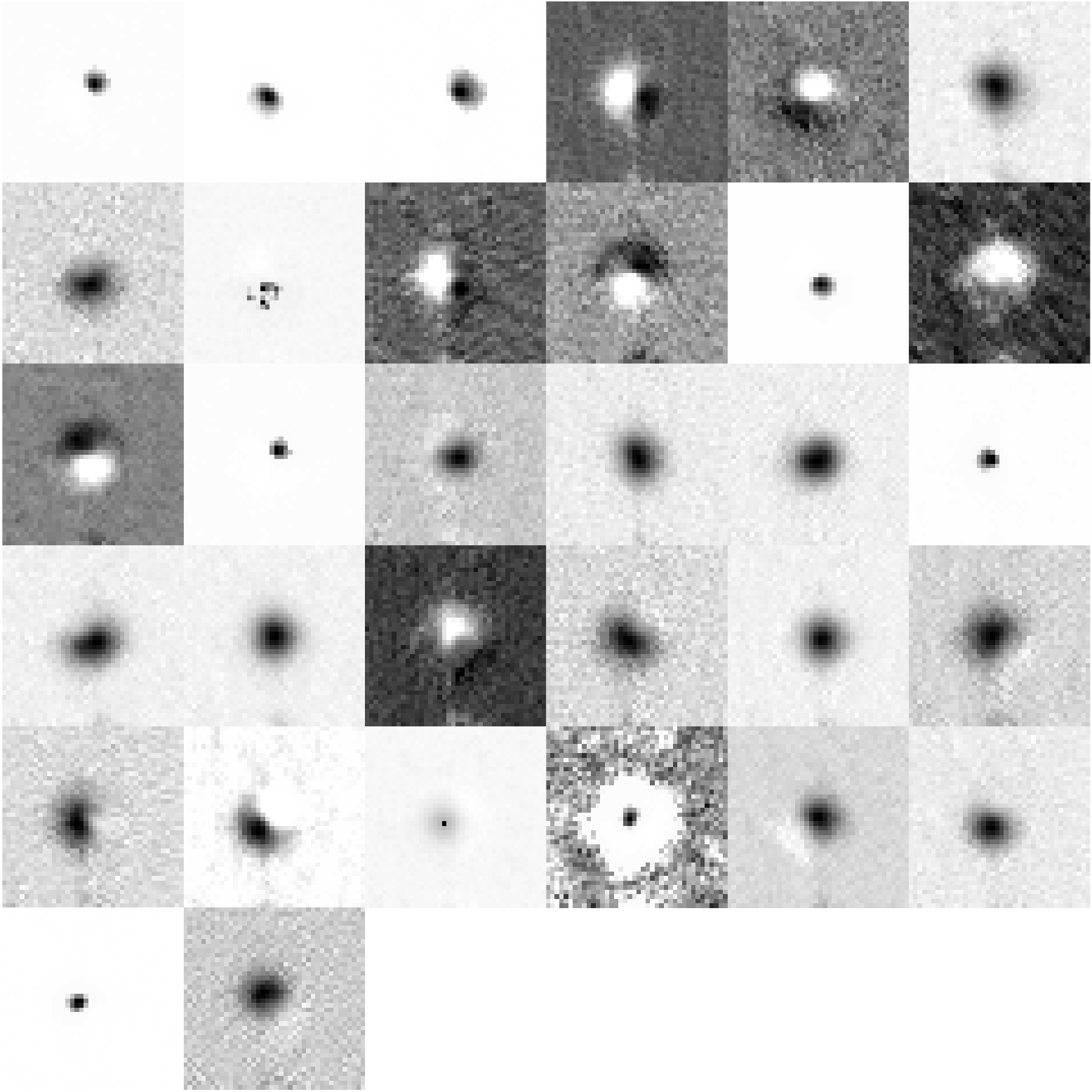}
    \caption{Layer 1 (32 maps$\times40\times40$)}
  \end{subfigure}
  \hfill
  \begin{subfigure}[c]{0.24\linewidth}
  \centering
    \includegraphics[width=\textwidth]{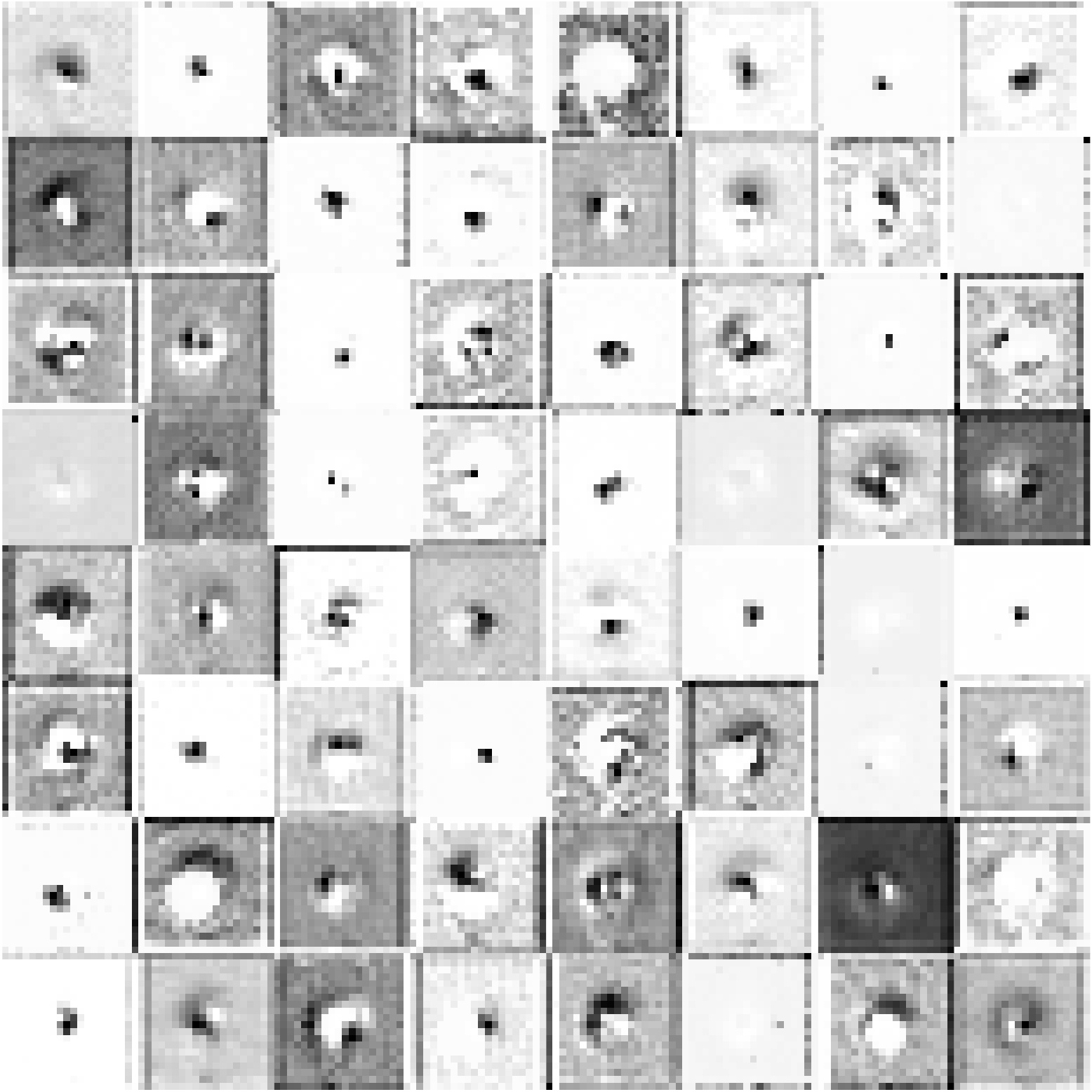}
    \caption{Layer 3 (64 maps$\times20\times20$)}
  \end{subfigure}
  \hfill
  \begin{subfigure}[c]{0.24\linewidth}
  \centering
    \includegraphics[width=\textwidth]{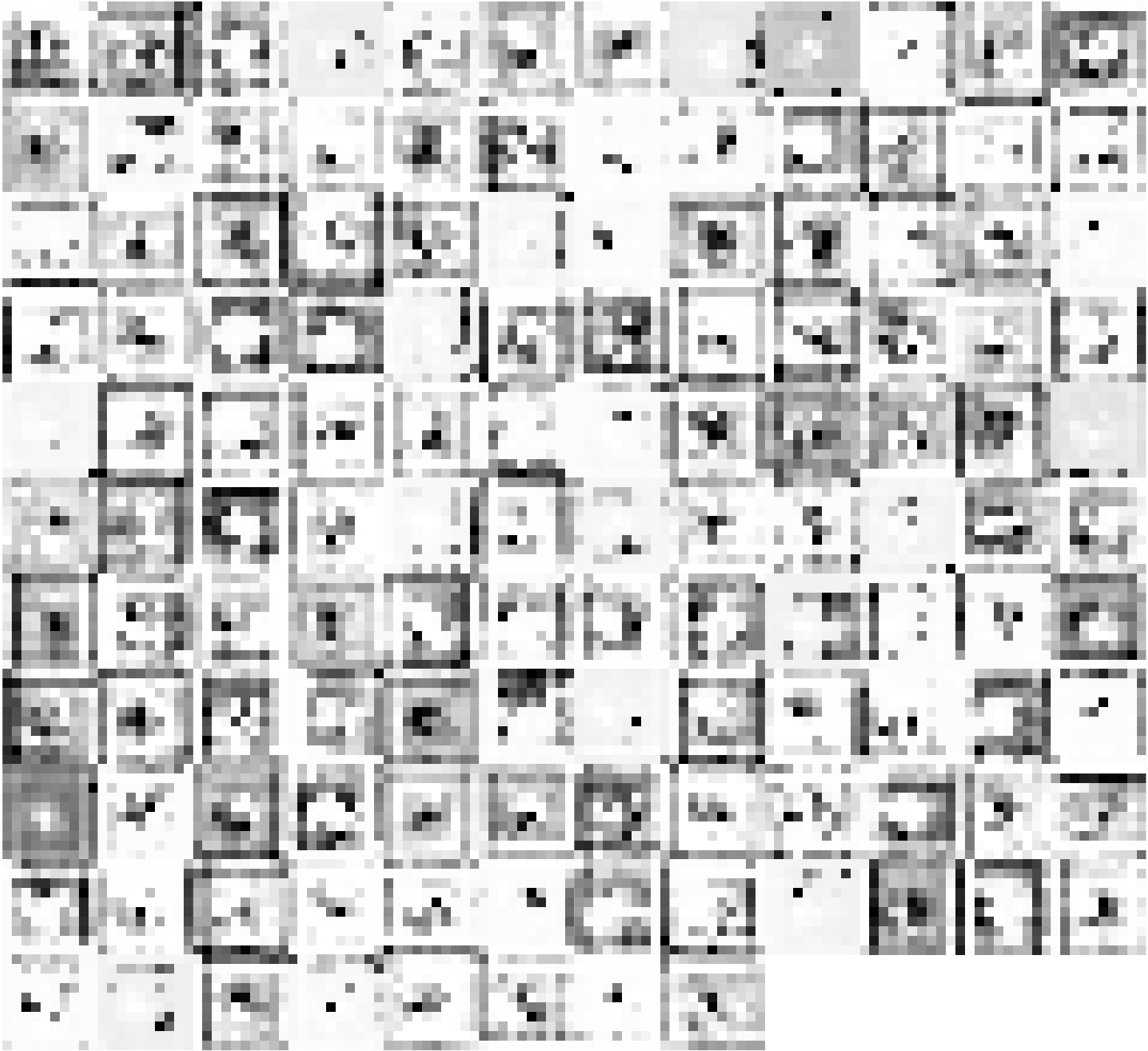}
    \caption{Layer 6 (128 maps$\times10\times10$)}
  \end{subfigure}
  \caption{
    Similar to Figure~\ref{fig:galaxy_activations} but for a star in the
    CFHTLenS data set.
    }
  \label{fig:star_activations}
\end{figure*}

As described in Section~\ref{sec:convnet},
we train our ConvNet model by monitoring its performance on the validation set.
Once we have finished training the model,
we evaluate its performance on the blind test set.
We also use the same training and validation sets to train
and tune the hyperparameters of 
$\rmn{TPC}_{\rmn{morph}}$ and $\rmn{TPC}_{\rmn{phot}}$, and
perform classifications on the same test set to compare their performance
with that of ConvNet.

We present in Table~\ref{table:clens_metrics} a summary of the results obtained
by applying ConvNet, $\rmn{TPC}_{\rmn{morph}}$, and $\rmn{TPC}_{\rmn{phot}}$
on the test set of the CFHTLenS data.
The bold entries highlight the best technique for any particular metric.
ConvNet outperforms $\rmn{TPC}_{\rmn{morph}}$ in
five
metrics
(AUC, $p_g$, CAL, $|\Delta N_g|/N_g$, 
and log loss
),
while $\rmn{TPC}_{\rmn{morph}}$ performs better in two metrics (MSE and $c_g$).
It is not surprising that $\rmn{TPC}_{\rmn{phot}}$, which is trained on only
magnitudes and colors, performs significantly worse than both ConvNet and
$\rmn{TPC}_{\rmn{phot}}$.
Thus, magnitudes and colors alone are not sufficient to separate stars from
galaxies,
and morphology is critical in separating stars from galaxies.
ConvNet is able to learn the morphological features automatically from
the images, and the performance of ConvNet is therefore comparable to
that of $\rmn{TPC}_{\rmn{morph}}$,
which is trained on both morphological and photometric attributes.

In Figure~\ref{fig:clens_mag}, we compare the galaxy purity
and star completeness values for ConvNet, $\rmn{TPC}_{\rmn{morph}}$, and
$\rmn{TPC}_{\rmn{phot}}$, as a function of  $i$-band magnitude
for the differential counts.
We use  kernel density estimation~\citep[KDE;][]{silverman1986density}
with a Gaussian kernel.
As the first panel shows, KDE is able to smooth the fluctuations in the distribution
without binning.
While ConvNet shows a slightly better performance than $\rmn{TPC}_{\rmn{morph}}$
in galaxy purity,
ConvNet performs slightly worse than $\rmn{TPC}_{\rmn{morph}}$ in star completeness.
Again, $\rmn{TPC}_{\rmn{phot}}$ performs significantly worse than both
ConvNet and $\rmn{TPC}_{\rmn{morph}}$, and this suggests that ConvNets are able to learn
the shape information automatically from the images.
We note that, at these operating conditions ($c_g=0.96$ or $p_s=0.97$),
both ConvNet and $\rmn{TPC}_{\rmn{morph}}$ outperform the
star-galaxy classification provided by
the CFHTLenS pipeline~\citep{hildebrandt2012cfhtlens} over all magnitudes.

In Figure~\ref{fig:clens_integrated}, we show the overall galaxy purity
and star completeness values as a function of $i$-band magnitude for the
integrated counts.
ConvNet is able to maintain a galaxy purity of 0.9972 up to $i\sim24.5$,
while the galaxy purity of $\rmn{TPC}_{\rmn{morph}}$ drops to 0.9963.
However, $\rmn{TPC}_{\rmn{morph}}$ performs better than ConvNet
in terms of star completeness, maintaining a purity of 0.9252 up to
$i\sim24.5$, while ConvNet drops to 0.8966.

We also show the galaxy purity and star completeness values as functions of
$g-r$ color in Figure~\ref{fig:clens_g_r}.
$\rmn{TPC}_{\rmn{morph}}$ provides slightly better
completeness and purity than ConvNet between $0.8 \la g-r \la 1.6$ while
ConvNet outperforms $\rmn{TPC}_{\rmn{morph}}$ in the remaining regions.

Figure~\ref{fig:clens_calibration} shows the calibration curves that compare
$P_{\rmn{gal}}$,
the fraction of objects that are galaxies (as determined from their spectra),
to $P_{\rmn{class}}$, the probabilistic outputs
produced by ConvNet and $\rmn{TPC}_{\rmn{morph}}$.
The calibration curve for our ConvNet model is nearly diagonal,
which implies that ConvNet is well-calibrated and we can treat its probabilistic
output as the probability that an object is a galaxy.
In contrast, the calibration curve
for the probabilistic output of $\rmn{TPC}_{\rmn{morph}}$
is apparently not as well-calibrated as ConvNet.
These calibration curves visually confirm the results in
Table~\ref{table:clens_metrics} that the calibration error
of ConvNet is about 20\% lower than that of $\rmn{TPC}_{\rmn{morph}}$.
While probabilistic predictions can be further calibrated by using,
\eg isotonic calibration~\citep{zadrozny2001obtaining},
we do not consider additional probability calibration in this work.

It is informative to visualize how an input image activates the neurons
in the convolutional layers.
Figures~\ref{fig:galaxy_activations} and \ref{fig:star_activations}
show the activations of the network when images of a galaxy and a star
are fed into the network.
The size of feature maps decreases with depth, and layers near the
input layer have fewer filters while the later layers have more.
The low-level features, \eg edges or blobs, of the input images are still
recognizable in the first convolutional layer.
Subsequent layers use these low-level features to detect higher-level features,
and the final layer is a classifier that uses these high-level features.
Thus, by performing hierarchical abstraction from low-level to high-level
features, ConvNets are able to utilize shape information in the classification
process.

\subsection{SDSS}
  \label{sec:results_sdss}

\begin{table*}
  \caption{
    A summary of the classification performance metrics
    as applied to the SDSS data.
    To obtain a galaxy completeness of $c_g=0.96$, we choose the threshold values
    0.7558, 0.9989, and 0.9360 for ConvNet, TPC${}_{\rmn{morph}}$, and
    TPC${}_{\rmn{phot}}$, respectively;
    for star purity $p_s=0.97$, we choose 0.6046, 0.0547, and 0.7449 for
    ConvNet, TPC${}_{\rmn{morph}}$, and TPC${}_{\rmn{phot}}$, respectively.
  }
  \centering
  \begin{tabular}{l c c c c c c c}
    \hline
    classifier & AUC & MSE & $p_{g}(c_g=0.96) $ & $c_{s}(p_s=0.97)$ & CAL & $ |\Delta N_g|/N_g$ &
log loss
\\
    \hline
    ConvNet               & 0.9952          & 0.0182          & 0.9915          & 0.9500          & \textbf{0.0243} & 0.0157 &
\textbf{0.0731}
\\
    TPC${}_{\rmn{morph}}$ & \textbf{0.9967} & \textbf{0.0099} & \textbf{0.9977} & \textbf{0.9810} & 0.0254          & \textbf{0.0044} & 
0.0914
\\
    TPC${}_{\rmn{phot}}$  & 0.9886          & 0.0283          & 0.9819          & 0.8879          & 0.0316          & 0.0160 & 
0.1372
\\
    \hline
  \end{tabular}
  \label{table:sdss_metrics}
\end{table*}

\begin{figure}
  \centering
  \includegraphics[width=\columnwidth]{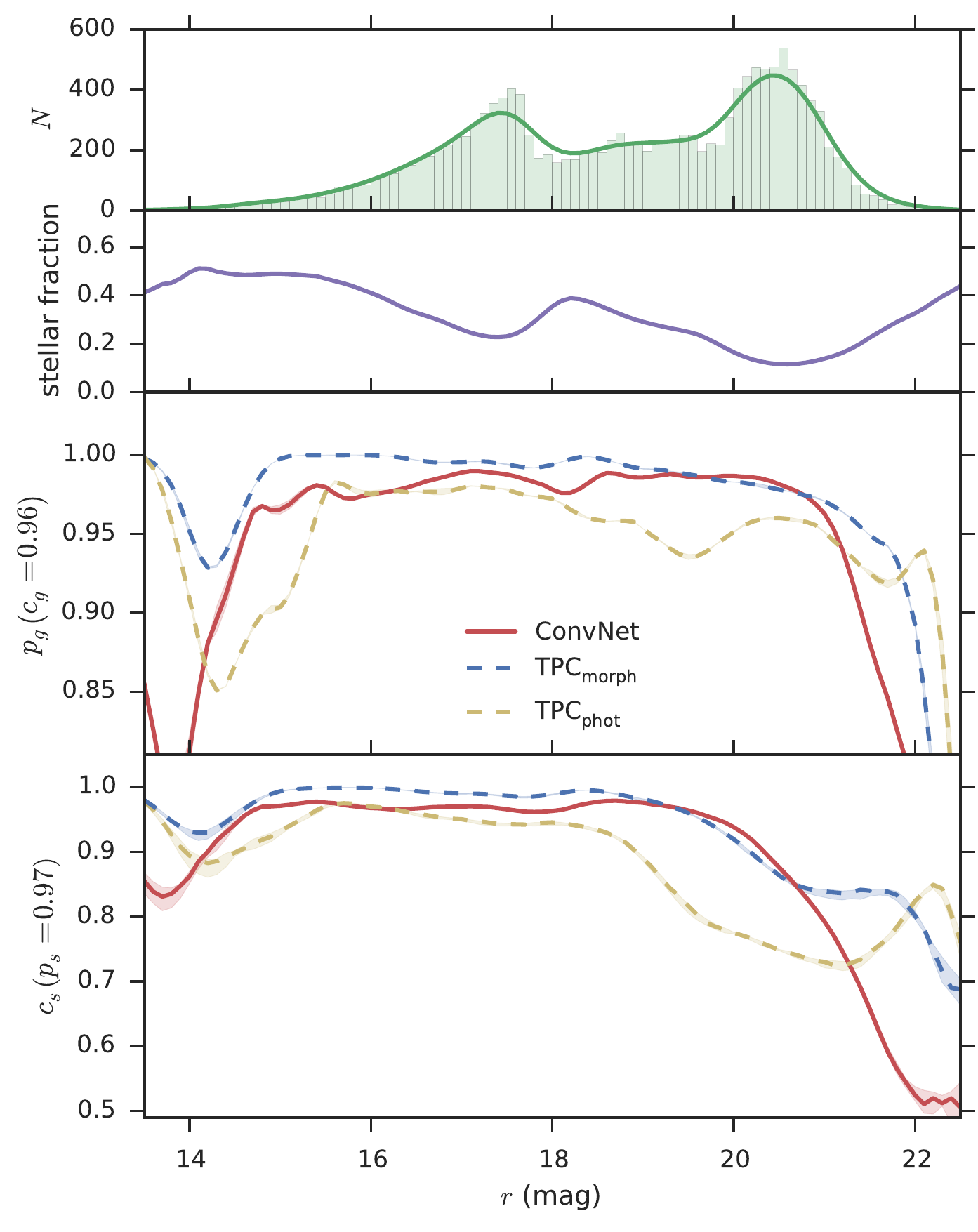}
  \caption{Galaxy purity and star completeness as function of the $r$-band
    magnitude for the differential counts in the SDSS data set.}
  \label{fig:sdss_mag}
\end{figure}

\begin{figure}
  \centering
  \includegraphics[width=\columnwidth]{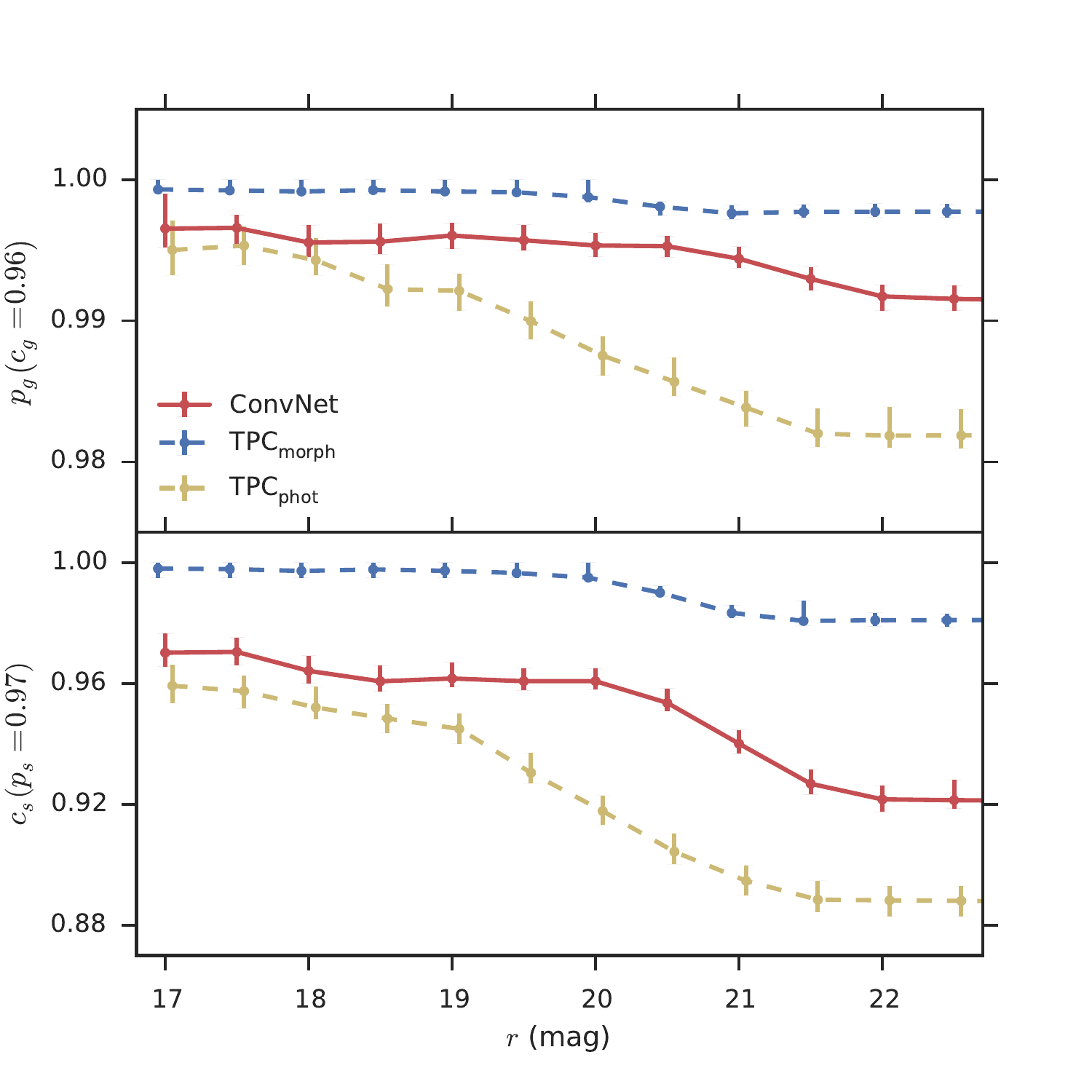}
  \caption{Galaxy purity and star completeness as functions of the $r$-band
    magnitude for the integrated counts in the SDSS data set.}
  \label{fig:sdss_integrated}
\end{figure}

\begin{figure}
  \centering
  \includegraphics[width=\columnwidth]{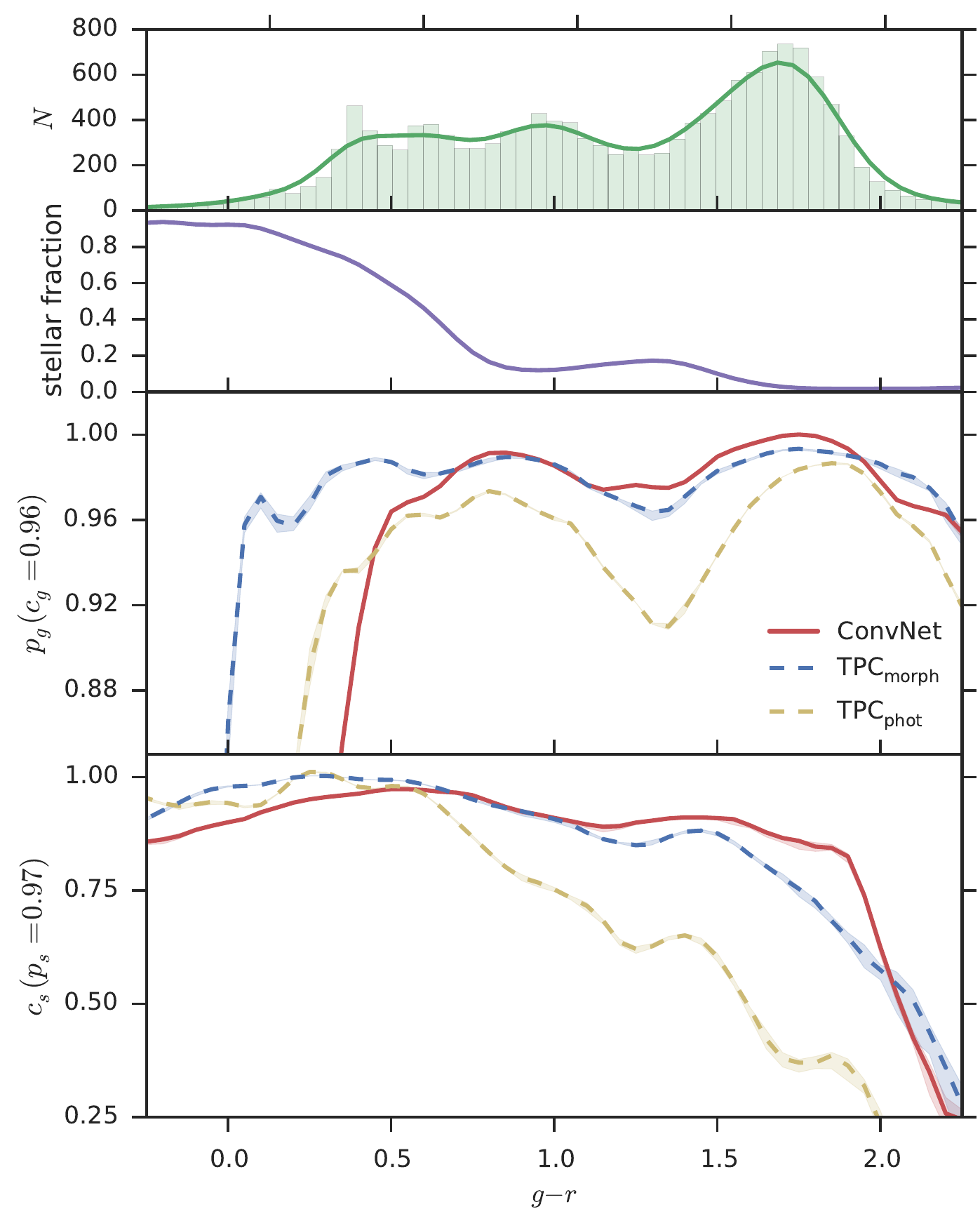}
  \caption{Similar to Figure~\ref{fig:sdss_mag} but as a function of
    $g-r$ color. The bin size of histogram in the top panel is 0.05.}
  \label{fig:sdss_g_r}
\end{figure}

\begin{figure}
  \centering
  \includegraphics[width=\columnwidth]{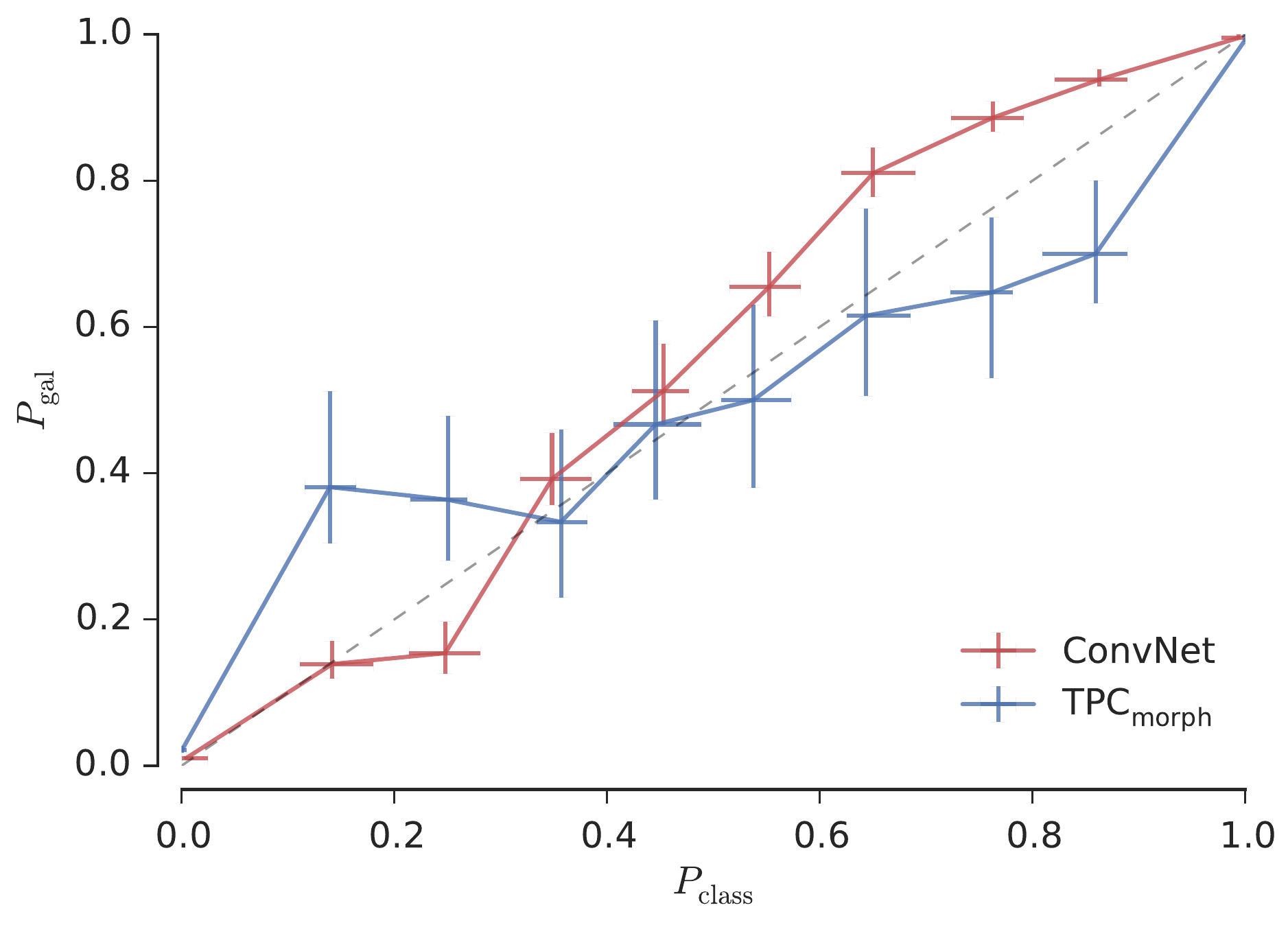}
  \caption{
    Calibration curves for ConvNet (red) and $\rmn{TPC}_{\rmn{morph}}$
    (blue) as applied to the SDSS data set.
    }
  \label{fig:sdss_calibration}
\end{figure}

\begin{figure*}
  \centering
  \begin{subfigure}[c]{0.24\linewidth}
  \centering
    \includegraphics[width=\textwidth]{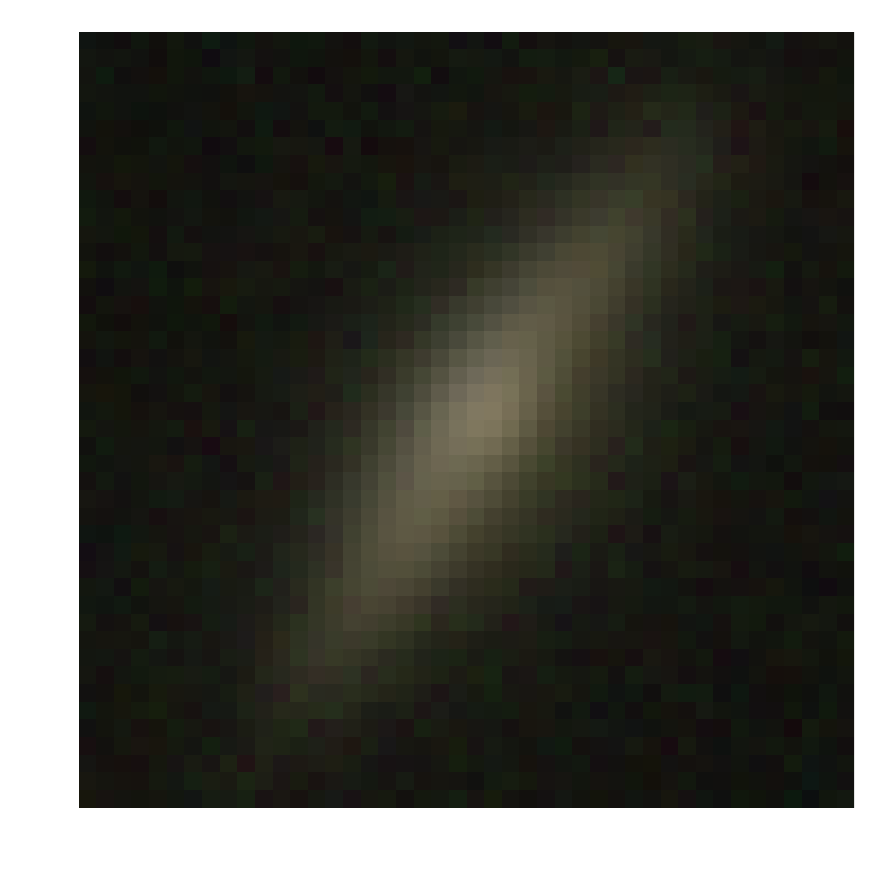}
    \caption{Input (5 bands$\times44\times44$)}
  \end{subfigure}
  \hfill
  \begin{subfigure}[c]{0.24\linewidth}
  \centering
    \includegraphics[width=\textwidth]{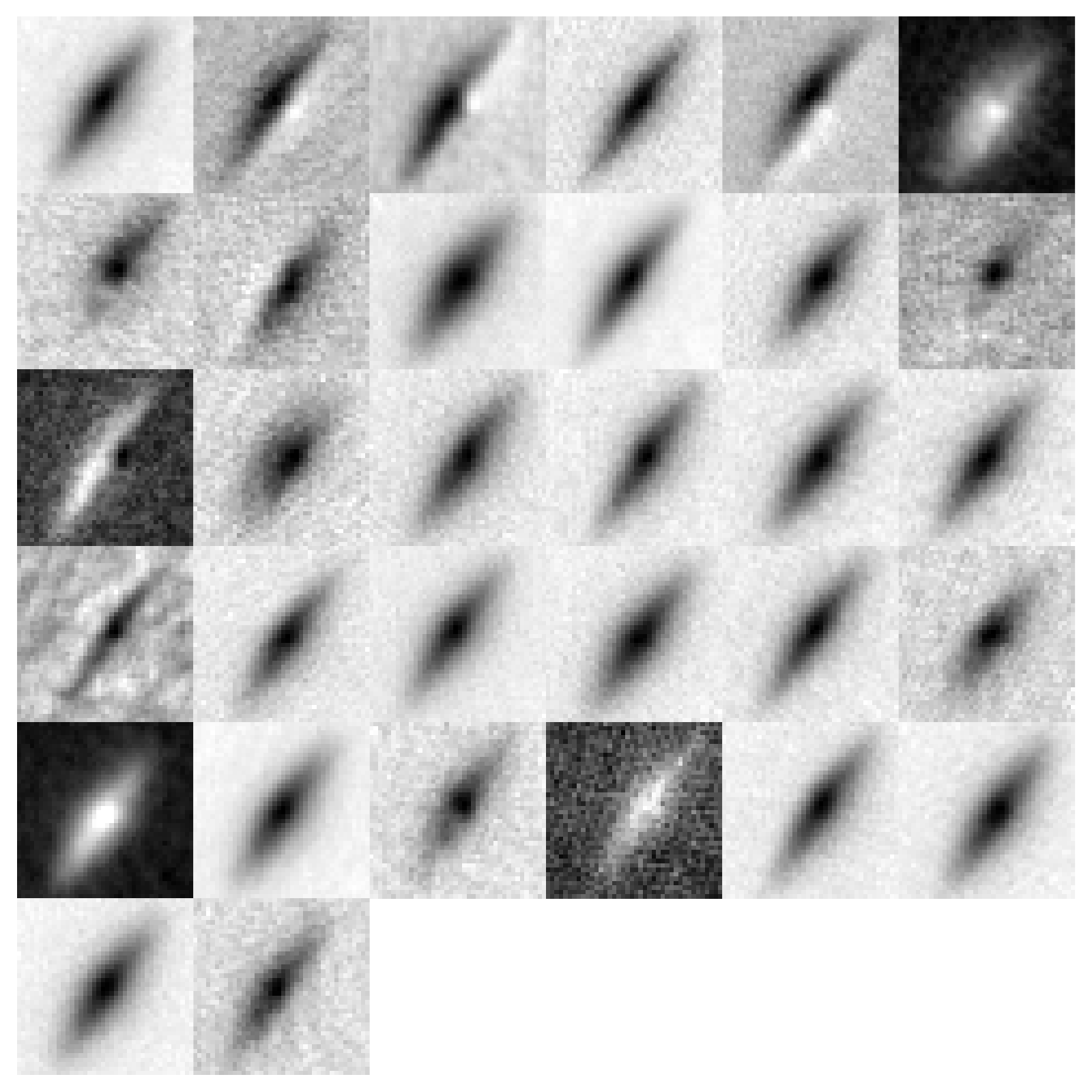}
    \caption{Layer 1 (32 maps$\times40\times40$)}
  \end{subfigure}
  \hfill
  \begin{subfigure}[c]{0.24\linewidth}
  \centering
    \includegraphics[width=\textwidth]{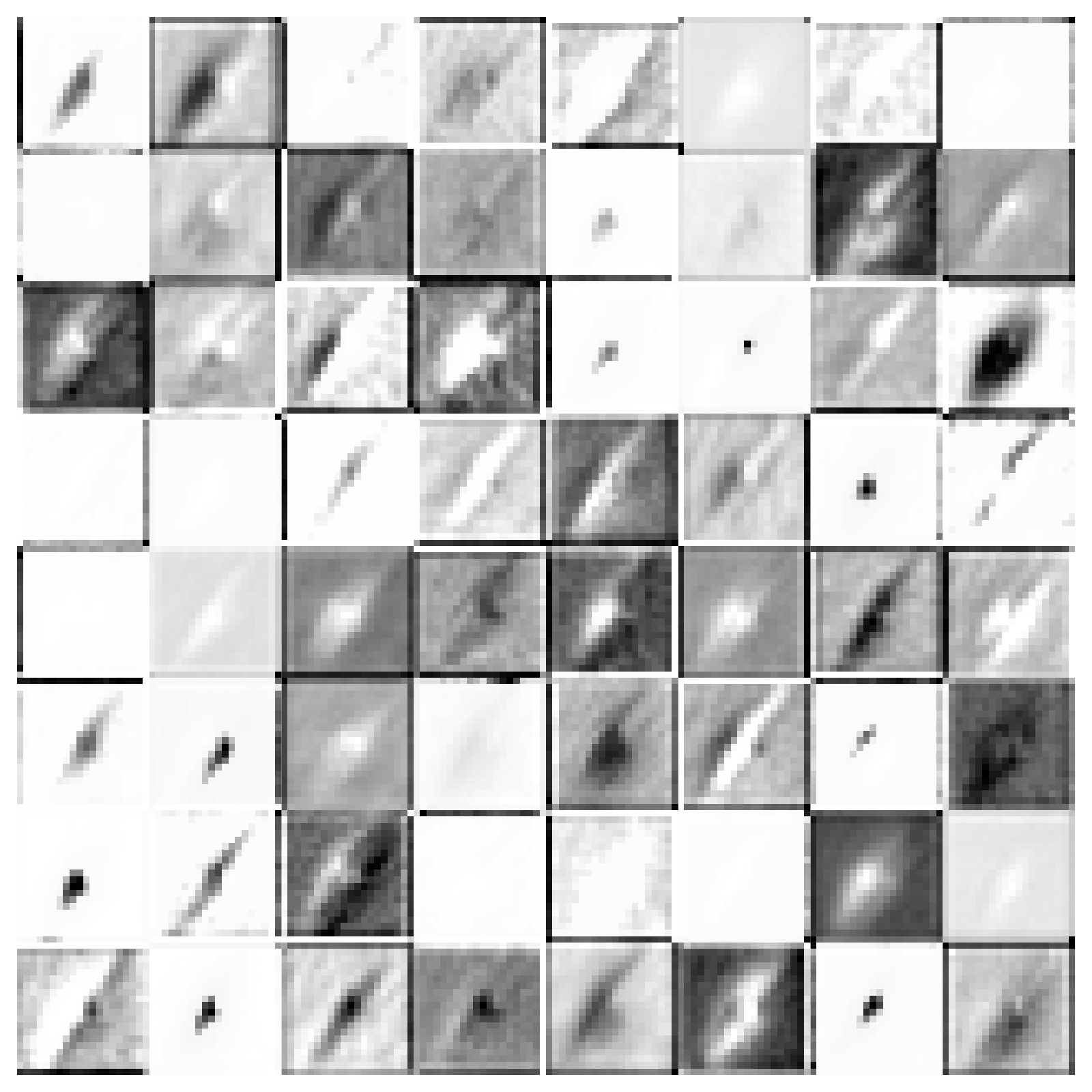}
    \caption{Layer 3 (64 maps$\times20\times20$)}
  \end{subfigure}
  \hfill
  \begin{subfigure}[c]{0.24\linewidth}
  \centering
    \includegraphics[width=\textwidth]{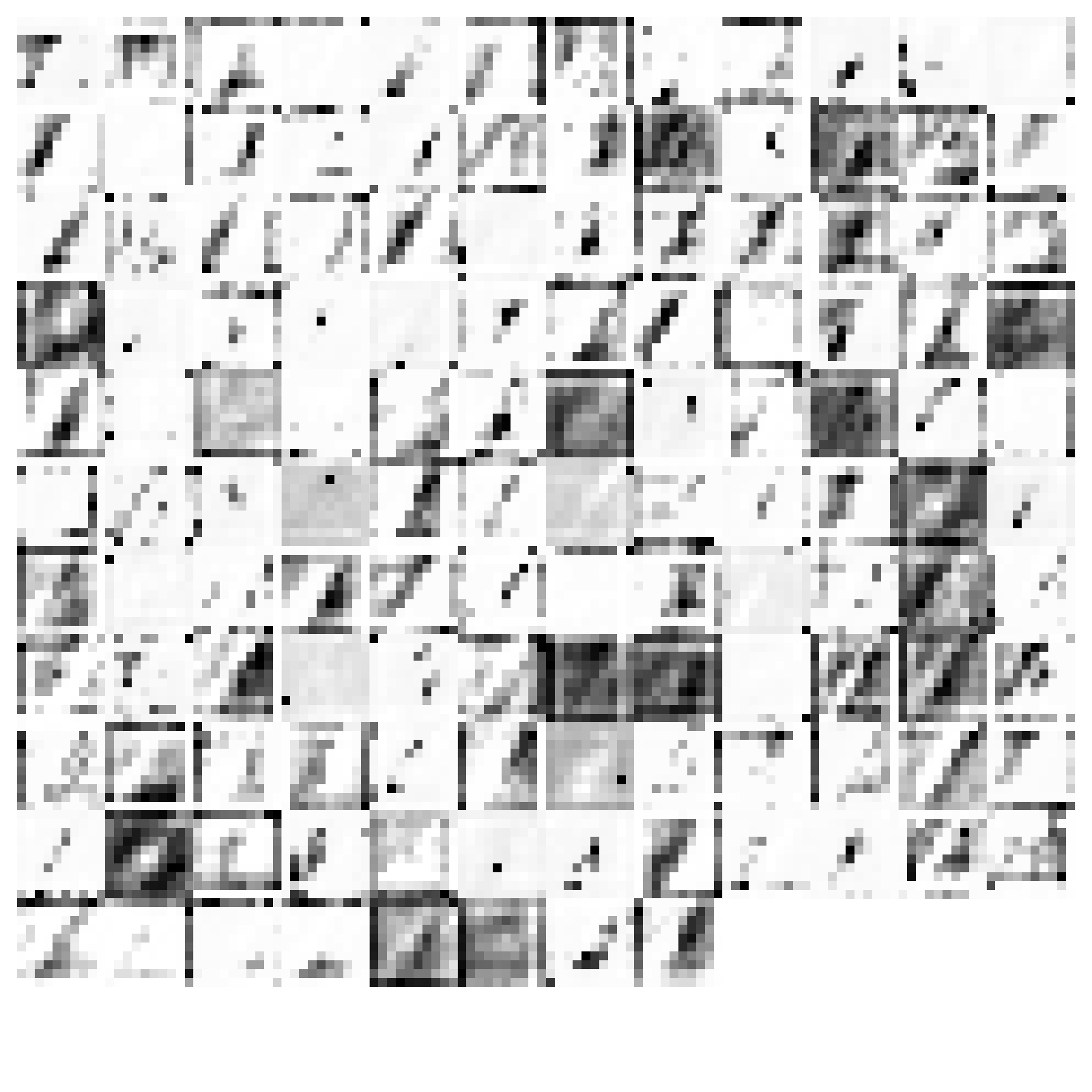}
    \caption{Layer 6 (128 maps$\times10\times10$)}
  \end{subfigure}
  \caption{
    Similar to Figure~\ref{fig:galaxy_activations} but for a galaxy in the
    SDSS data set.
    }
  \label{fig:sdss_galaxy_activations}
\end{figure*}

\begin{figure*}
  \centering
  \begin{subfigure}[c]{0.24\linewidth}
  \centering
    \includegraphics[width=\textwidth]{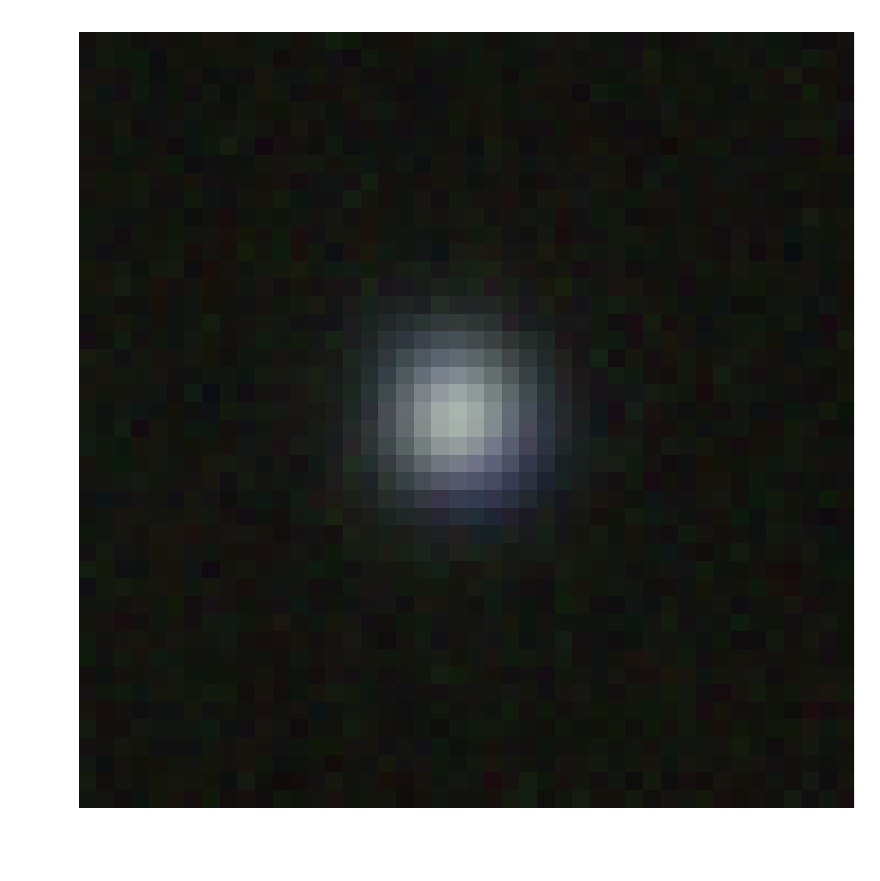}
    \caption{Input (5 bands$\times44\times44$)}
  \end{subfigure}
  \hfill
  \begin{subfigure}[c]{0.24\linewidth}
  \centering
    \includegraphics[width=\textwidth]{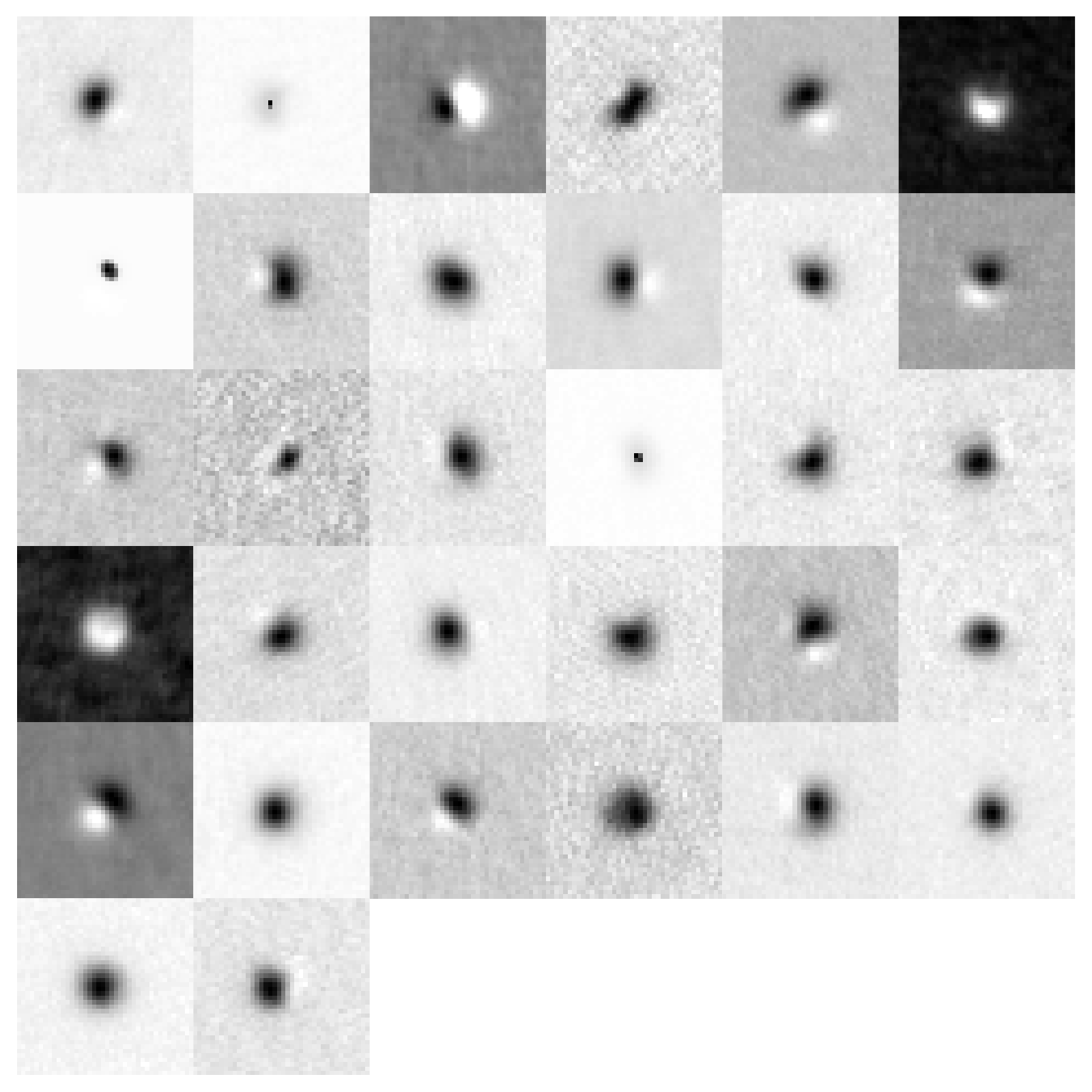}
    \caption{Layer 1 (32 maps$\times40\times40$)}
  \end{subfigure}
  \hfill
  \begin{subfigure}[c]{0.24\linewidth}
  \centering
    \includegraphics[width=\textwidth]{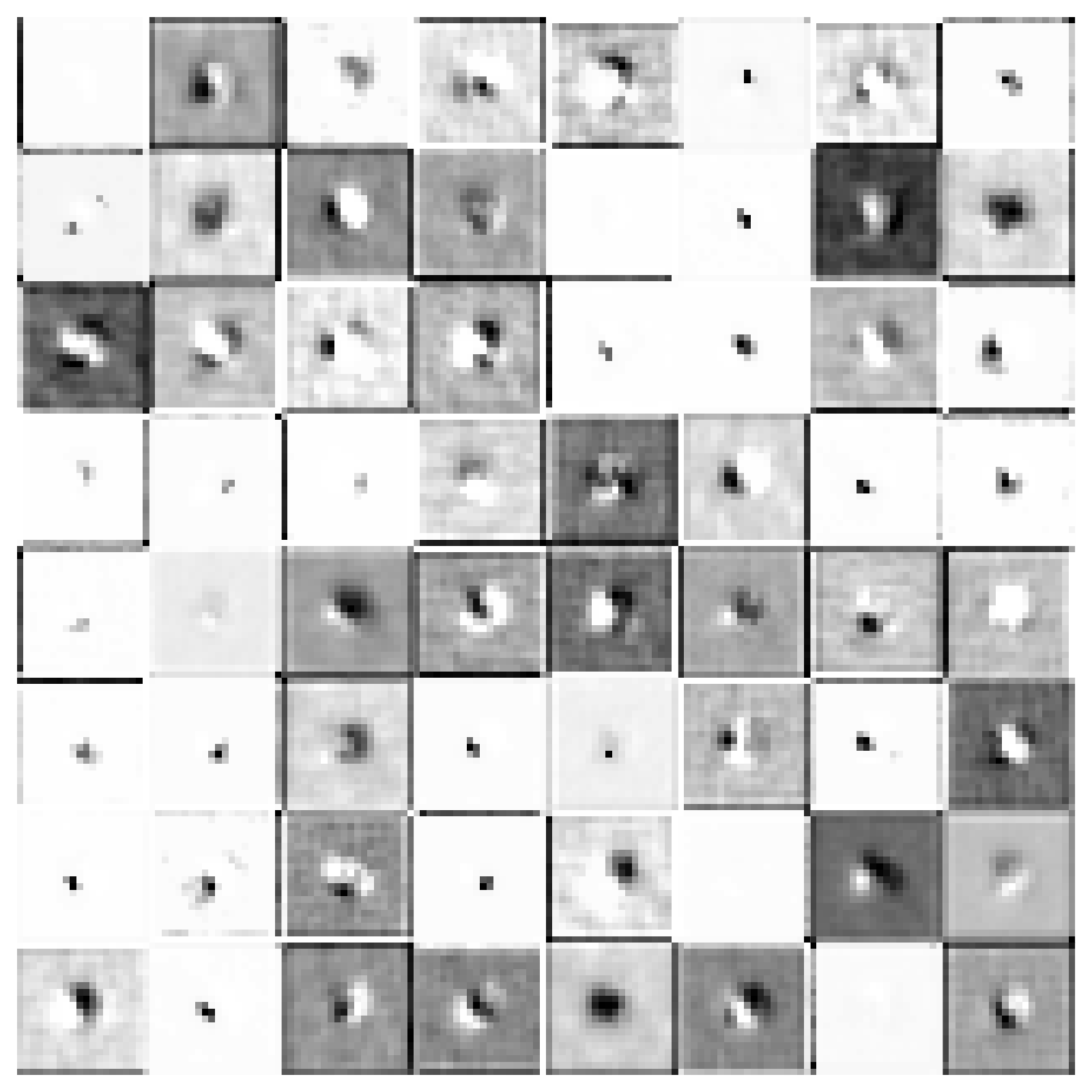}
    \caption{Layer 3 (64 maps$\times20\times20$)}
  \end{subfigure}
  \hfill
  \begin{subfigure}[c]{0.24\linewidth}
  \centering
    \includegraphics[width=\textwidth]{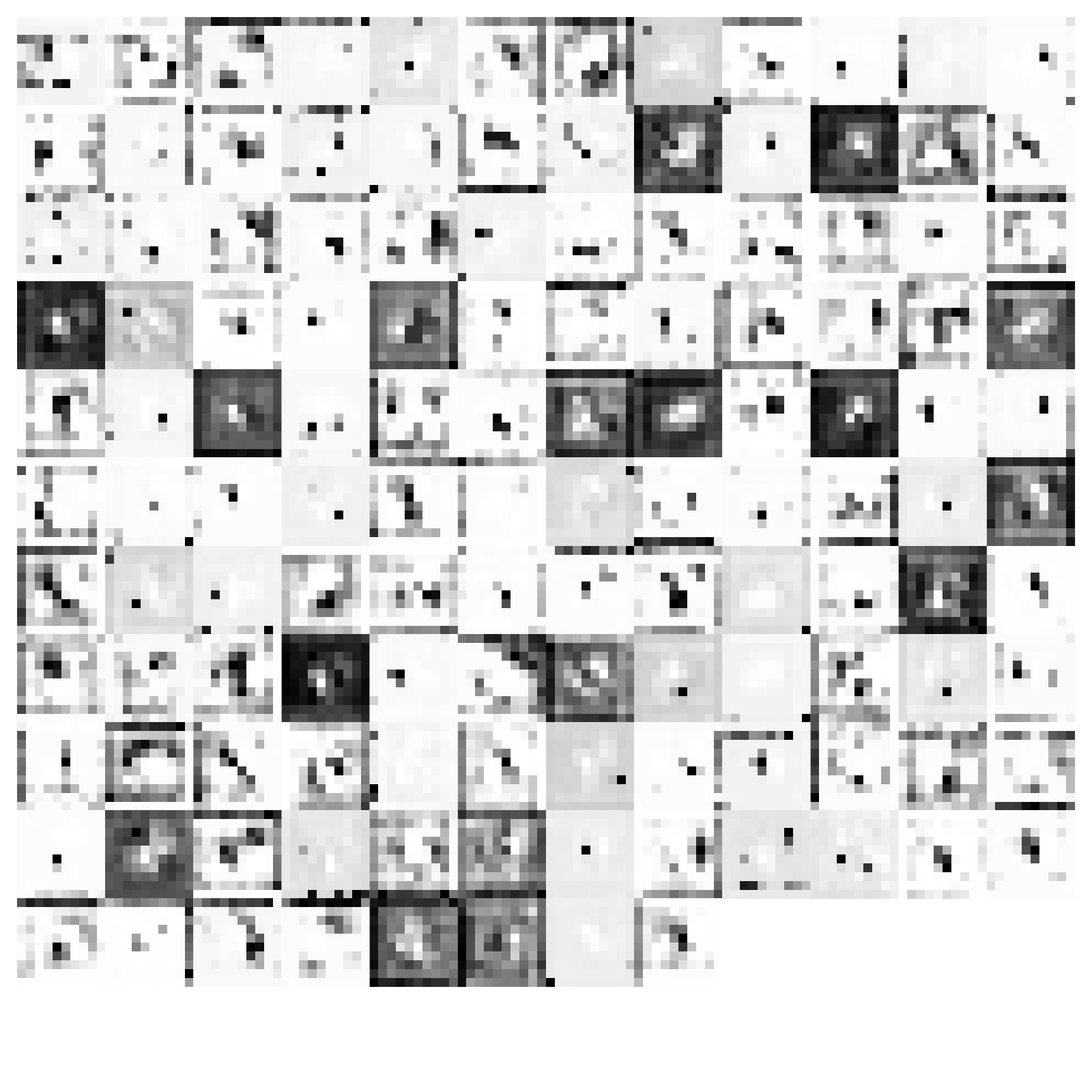}
    \caption{Layer 6 (128 maps$\times10\times10$)}
  \end{subfigure}
  \caption{
    Similar to Figure~\ref{fig:sdss_galaxy_activations} but for a star in the
    SDSS data set.
    }
  \label{fig:sdss_star_activations}
\end{figure*}

We have also trained and tested our ConvNet model on the SDSS data set, and
we present in Table~\ref{table:sdss_metrics} the same six metrics for ConvNet,
$\rmn{TPC}_{\rmn{morph}}$, and $\rmn{TPC}_{\rmn{phot}}$.
The bold entries highlight the best technique for any particular metrics.
In contrast with the CFHTLenS data set in Section~\ref{sec:results_cfht},
it is apparent that $\rmn{TPC}_{\rmn{morph}}$ outperforms ConvNet
in all metrics except $\rmn{CAL}$
and cross-entropy.
Both ConvNet and $\rmn{TPC}_{\rmn{morph}}$ still outperform
$\rmn{TPC}_{\rmn{phot}}$ in all six metrics by a significant amount, as
magnitudes and colors alone are not sufficient to separate stars from galaxies.
Although ConvNet performs worse than $\rmn{TPC}_{\rmn{morph}}$ on the SDSS
data, its performance is much closer to $\rmn{TPC}_{\rmn{morph}}$, as ConvNet
is able to learn the shape information automatically from the images.

In Figure~\ref{fig:sdss_mag}, we compare the galaxy purity
and star completeness values for ConvNet, $\rmn{TPC}_{\rmn{morph}}$, and
$\rmn{TPC}_{\rmn{phot}}$ as a
function of $r$-band magnitude for the differential counts in the SDSS data.
We note that $\rmn{TPC}_{\rmn{morph}}$ outperforms the star-galaxy
classifier used by the SDSS pipeline (\ie an object is classified as a galaxy
if $\rmn{concentration} > 0.145$) over all magnitudes.
We do not show the SDSS classifications to avoid cluttering the plots.
While ConvNet shows a similar but slightly worse performance than
$\rmn{TPC}_{\rmn{morph}}$,
the galaxy purity and star completeness values of ConvNet begin to
drop at faint magnitudes $i \la 21$.
Again, $\rmn{TPC}_{\rmn{phot}}$ performs significantly worse than both
ConvNet and TPC at bright magnitudes.
One reason that ConvNet fails to outperform $\rmn{TPC}_{\rmn{phot}}$,
especially at faint magnitudes,
might be its over-reliance on morphological features.
Near a survey's limit, the measurement uncertainties generally increase,
and morphology is not a reliable metric for star-galaxy classification.
Another possibility is that data augmentation has a negative effect at faint
magnitudes, as the network may get confused by additional examples of faint
galaxies that look like point sources.
Data augmentation however is indispensable, since it improves the overall
performance greatly.

In Figure~\ref{fig:sdss_integrated}, we show the overall galaxy purity
and star completeness values as a function of magnitude for the integrated
counts.
ConvNet is able to maintain a galaxy purity of 0.9915 up to $i\sim22.5$,
while $\rmn{TPC}_{\rmn{morph}}$ provides a galaxy purity of 0.9977.
$\rmn{TPC}_{\rmn{morph}}$ also outperforms ConvNet
in terms of star completeness, maintaining a purity of 0.9810 up to
$i\sim22.5$, while the star completeness of ConvNet drops to 0.9500.

We also show the galaxy purity and star completeness values as a function of
$g-r$ color in Figure~\ref{fig:sdss_g_r}.
ConvNet performs slightly better than $\rmn{TPC}_{\rmn{morph}}$ in both
galaxy completeness and star purity between $0.7 \la g-r \la 2.0$,
where the stellar fraction is relatively low.
On the other hand, both $\rmn{TPC}_{\rmn{morph}}$ and $\rmn{TPC}_{\rmn{phot}}$ outperform
ConvNet in the region $g-r \la 0.8$ where the stellar fraction is higher.

Figure~\ref{fig:sdss_calibration} shows
the calibration curves of ConvNet and $\rmn{TPC}_{\rmn{morph}}$.
The calibration curve of ConvNet in Figure~\ref{fig:sdss_calibration}
is not as well-calibrated as the calibration curve in
Figure~\ref{fig:clens_calibration},
where the same ConvNet model was applied to the CFHTLenS data set.
However, ConvNet may still be better calibrated than $\rmn{TPC}_{\rmn{morph}}$,
even when it is applied to the SDSS data set.
Although it is not straightforward to compare the two calibration curves by
visual inspection, Table~\ref{table:sdss_metrics} shows that the CAL metric of
ConvNet is lower than that of $\rmn{TPC}_{\rmn{morph}}$.

Figures~\ref{fig:sdss_galaxy_activations} and \ref{fig:sdss_star_activations}
show the activations when images of a galaxy and a star are fed into the network. 
Similarly to Figures~\ref{fig:galaxy_activations} and \ref{fig:star_activations}
in Section~\ref{sec:results_cfht},
the feature maps show hierarchical abstraction from low-level features in the
first convolutional layer to high-level features in the subsequent layers.
This hierarchical abstraction is what enables ConvNets to learn morphological
features automatically from images.


\section{Conclusions}
  \label{sec:conclusions}

We have presented a convolutional neural network for classifying stars and
galaxies in the SDSS and CFHTLenS photometric images.
For the CFHTLenS data set, the network is able to provide a classification that
is as accurate as a random forest algorithm (TPC), while the probability estimates of
our ConvNet model appear to be better calibrated.
When the same network architecture is applied to the SDSS data set,
the network fails to outperform TPC,
but the probabilities are still slightly better calibrated.
The major advantage of ConvNets is that useful features are learned
automatically from images, while traditional machine learning
algorithms require feature engineering as a separate process
to produce accurate classifications.

ConvNets have recently achieved record-breaking results in many image
classification tasks~\citep{lecun2015deep} and have been quickly and widely
adopted by the computer vision community.
One of the main reasons for the success is that ConvNets are general-purpose
algorithms that are applicable to a variety of problems without the need for
designing a feature extractor.
The lack of requirement for feature extraction is a huge advantage, \eg
when the task is to classify 1,000 classes in the ImageNet data
set~\citep{russakovsky2015imagenet},
as a good feature extractor for identifying images of cats would be of little use
for classifying sailboats, and it is impractical to design a separate
feature extractor for each class.
However, when there already exists a good feature extractor for the problem
at hand, \eg the concentration parameter, the weight-averaged
\texttt{spread\_model} parameter from the Dark Energy
Survey~\citep{desai2012blanco,crocce2016galaxy},
or even the \texttt{SExtractor} software,
conventional machine learning algorithms that have been shown to be effective,
such as TPC~\citep{carrascokind2013tpz, kim2015hybrid}, remain a viable option.
As the ``no free lunch'' theorem~\citep{wolpert1996lack} states,
there is no one model that works for every problem.
For the CFHTLenS data set, our ConvNet model outperforms TPC.
Since the SDSS catalog provides the concentration parameter that is highly
optimized for star-galaxy classification, TPC works better for SDSS.

Although we used various techniques to combat overfitting,
it is possible that our ConvNet model has overfit the data.
Overfitting could explain why our ConvNet model with maximal information
fails to significantly outperform a standard machine learning algorithm
that uses the reduced summary information from catalogs.
The most effective way to prevent overfitting would be to simply collect more training images
with spectroscopic follow-up, as the performance of ConvNets generally
improves with more training data.
However, spectroscopic observations are expensive and time-consuming,
and it is unclear if sufficient training data will be available
in future photometric surveys.
If enough training data become available in DES or LSST, ConvNets
become an attractive option because it can be applied immediately on reduced, calibrated images
to produce well-calibrated posterior probabilities.
We also note that using more training images will require the use of multi-GPU systems,
which was beyond the scope of the present work.

Deep learning is a rapidly developing field, and recent developments include
improved network architectures.
For future work, we plan to train more ConvNet variants, such as the
Inception Module~\citep{szegedy2015going} and Residual Network~\citep{he2015deep}.
To improve the predictive performance,
we have combined the predictions of different models across multiple
transformations of the input images (Section~\ref{sec:bmc}).
To further improve the performance, we could also train several networks
with different architectures and combine the models.
For example, the winning solution of \cite{dieleman2015rotation}
for the Galaxy Zoo challenge was based on a ConvNet model,
and it required averaging many sets of predictions from models with different
neural network architectures.
Futhermore, future work could compare the performance of other deep learning
variants, such as deep belief networks~\citep{hinton2006fast},
deep Boltzmann machines~\citep{salakhutdinov2009deep}, or
multilayer perceptrons~\citep{wasserman1988neural}.

It is also likely that the performance will be improved
not only by training multiple network architectures,
but also by combining them with different star-galaxy classifiers.
In \citet{kim2015hybrid}, we combined a purely morphological classifier,
a supervised machine learning method (TPC),
an unsupervised machine learning method based on self-organizing maps,
and a hierarchical Bayesian template fitting method, and
demonstrated that our combination technique improves the overall
performance over any individual classification method.
ConvNets could be included as a different machine learning paradigm in this
classifier combination framework to produce further improvements in
predictive performance.

Our ConvNet model is a supervised algorithm, and one of the criticisms
of supervised techniques is their difficulty in extrapolating past the limits
of available spectroscopic training data.
Since it is difficult to assess the classification performance without a deeper
spectroscopic sample, we evaluated the performance using a test set
that is drawn from the same underlying distribution as the spectroscopic sample.
However, when our ConvNet model
---trained on sources from a spectroscopic sample---
is applied to a photometric sample
---which is often fainter than our training set--- 
the performance of ConvNet will be less reliable.
Combining our ConvNet model with unsupervised methods (\eg a template fitting
method) in the aforementioned meta-classification framework will
help improve the efficacy of star-galaxy classification
beyond the limits of spectroscopic training data.

Finally, we are currently exploring different strategies for including objects
that are neither stars nor galaxies (\eg quasars).
The results of this multi-class problem will be presented in subsequent papers.


\section*{Acknowledgements}

The authors thank the referee for a careful reading of the manuscript
and comments that improved this work.
We acknowledge support from the 
National Science Foundation Grant No.\ AST-1313415.
RJB acknowledges support as an Associate
within the Center for Advanced Study at the University of Illinois.

This work used the Extreme Science and Engineering Discovery Environment
(XSEDE), which is supported by National Science Foundation grant number
ACI-1053575.

This work has made use of the Theano library
(http://deeplearning.net/software/theano/),
the Lasagne library
(http://lasagne.readthedocs.io/),
and the Nolearn library
(http://pythonhosted.org/nolearn/).

This work is based on observations obtained with MegaPrime/MegaCam, a
joint project of CFHT and CEA/DAPNIA, at the Canada-France-Hawaii
Telescope (CFHT) which is operated by the National Research Council
(NRC) of Canada, the Institut National des Sciences de l'Univers of
the Centre National de la Recherche Scientifique (CNRS) of France, and
the University of Hawaii. This research used the facilities of the
Canadian Astronomy Data Centre operated by the National Research
Council of Canada with the support of the Canadian Space Agency.
CFHTLenS data processing was made possible thanks to significant
computing support from the NSERC Research Tools and Instruments grant
program.

Funding for SDSS-III has been provided by the Alfred P. Sloan Foundation, the
Participating Institutions, the National Science Foundation, and the U.S.
Department of Energy Office of Science. The SDSS-III web site is
http://www.sdss3.org/.

SDSS-III is managed by the Astrophysical Research Consortium for the
Participating Institutions of the SDSS-III Collaboration including the
University of Arizona, the Brazilian Participation Group, Brookhaven National
Laboratory, Carnegie Mellon University, University of Florida, the French
Participation Group, the German Participation Group, Harvard University, the
Instituto de Astrofisica de Canarias, the Michigan State/Notre Dame/JINA
Participation Group, Johns Hopkins University, Lawrence Berkeley National
Laboratory, Max Planck Institute for Astrophysics, Max Planck Institute for
Extraterrestrial Physics, New Mexico State University, New York University,
Ohio State University, Pennsylvania State University, University of Portsmouth,
Princeton University, the Spanish Participation Group, University of Tokyo,
University of Utah, Vanderbilt University, University of Virginia, University
of Washington, and Yale University.

\bibliographystyle{mnras}
\bibliography{main}

\bsp	
\label{lastpage}
\end{document}